\documentclass[12pt]{article}

\usepackage{latexsym,amsmath,amssymb,theorem,epsfig} 

\DeclareFontFamily{OT1}{rsfs10}{} 
\DeclareFontShape{OT1}{rsfs10}{m}{n}{ <-> rsfs10 }{} 
\DeclareMathAlphabet{\mathscript}{OT1}{rsfs10}{m}{n} 

\numberwithin{equation}{section}

\newcommand{\ns}{\normalsize}

\def\a{\alpha}
\def\b{\beta}
\def\g{\gamma}

\def\s{\sigma}

\def\cC{{\cal C}}
\def\cE{{\cal E}}

\def\cN{{\cal N}}
\def\cO{{\cal O}}

\def\cS{{\cal S}}

\theoremstyle{plain} 

{\theorembodyfont{\rmfamily} }

\begin{document}


\begin{titlepage}

\vspace{-5cm}

\title{
  \hfill{\ns }  \\[1em]
   {\LARGE Instanton Moduli in String Theory} 
\\[1em] } 
\author{
   Evgeny I. Buchbinder,$^{1}$ 
   Burt A. Ovrut$^{2}$ and 
   Rene Reinbacher$^{3}$
     \\[0.5em]
   {\ns $^1$School of Natural Sciences, Institute for Advanced Study} \\[-0.4cm]
{\ns Einstein Drive, Princeton, NJ 08540, USA}\\[0.2cm]
{\ns $^2$Department of Physics and Astronomy, University of 
Pennsylvania} \\[-0.4em]
   {\ns Philadelphia, PA 19104, USA}\\[0.2cm]
{\ns $^3$ Department of Mathematics, Harvard University, Cambridge, MA 02138, USA}\\}

\date{}

\maketitle

\begin{abstract}

Expressions for the number of moduli of arbitrary $SU(n)$ vector bundles 
constructed via Fourier-Mukai transforms of spectral data over Calabi-
Yau threefolds are derived and presented. This is done within the context
of simply connected, elliptic Calabi-Yau threefolds with base ${\mathbb F}_{r}$,
but the methods have wider applicability. The condition for a vector bundle to 
possess the minimal number of moduli for fixed $r$ and $n$ is discussed and an 
explicit formula for the minimal number of moduli is presented. In addition, 
transition moduli for small instanton phase transitions involving non-positive 
spectral covers are defined, enumerated and given a geometrical interpretation.

\end{abstract}

\thispagestyle{empty}

\end{titlepage}


\section{Introduction}


It was shown in~\cite{Witten96, AA, BB, CC, DD} 
that compactifying Ho\v rava-Witten theory~\cite{HW1, HW2}
on a Calabi-Yau threefold, $X$, with an appropriate holomorphic vector
bundle, $V$, can lead, on the observable brane, to realistic GUT and 
standard-like models at low energy. It is clear that in this context
instantons with the ``standard embedding'', that is, vector bundles 
where $V=TX$, do not play a special role and that one should consider 
more generalized vector bundles.  Based on the work in~\cite{FMW, Donagi}, 
large classes of 
such bundles were constructed. In~\cite{EE, DLOW}, stable holomorphic 
vector bundles 
with structure group $SU(n)$ over simply connected elliptic 
Calabi-Yau threefolds were
presented. These lead to GUT models in four-dimensions with 
gauge groups $SU(5)$
and $Spin(10)$, for example. 
In a series of papers~\cite{Standard1, Standard2, FF, 
Rene1, Rene2, SU(4), Volker}, 
stable holomorphic 
vector bundles with structure group $SU(n)$ were constructed, using new 
methods, over Calabi-Yau 
threefolds with non-trivial fundamental group. 
By combining these bundles with 
Wilson lines, one can produce quasi-realistic low energy models 
containing the standard model gauge group $SU(3)_{C} \times SU(2)_{L} 
\times U(1)_{Y}$. Various properties of these generalized heterotic 
vacua have been 
studied, including the moduli space and BPS 
spectrum~\cite{Five1, Five2, Lalak1, Deren, Guralnik} 
of the associated 
five-branes, small instanton phase 
transitions~\cite{Seiberg, Kachru, OPP, BDO1, Ruip, Reduce}, non-perturbative 
superpotentials~\cite{BBS, Wittensuper, Lima1, Lima2, Moore, BDO2, BDO3, Beasley},  
supersymmetry breaking~\cite{Horava, Break1, Break2, Lalak2, Lalak3, Nilles} and 
moduli stabilization~\cite{CurK, BO, Gukov, Becker, Raise}.
Recently, it was shown 
how to compute the particle spectrum produced by generalized vector 
bundles, both in GUT theories~\cite{Yang1, Yang2} 
and in theories of the standard model~\cite{Yang3}.
Such theories also impact early universe cosmology. In particular, they
underlie the formulation of Ekpyrotic cosmology~\cite{Justin1, Justin2, 
Justin3}.

An important aspect of generalized vector bundles is their moduli space. 
Vector bundle moduli appear as uncharged complex scalar fields in the low 
energy effective theory and, as such, strongly impact particle phenomenology 
and cosmology. The computation of the number and properties of these moduli 
is a non-trivial exercise in sheaf cohomology. This was carried out for generalized
bundles on simply connected elliptic Calabi-Yau threefolds in~\cite{BDO1}, under the simplifying
assumption that the spectral covers associated with these bundles are ``positive''. 
Within this context, the moduli associated with phase transitions, ``transition'' moduli,
were defined and studied~\cite{BDO1}. Furthermore, in~\cite{BDO2, BDO3} the contribution of vector bundle 
moduli to non-perturbative superpotentials was explicitly computed. Unfortunately, 
the number of positive spectral covers is only a small subset of all allowed covers.
In addition, one can show that most phenomenologically acceptable heterotic vacua 
correspond to spectral covers that are explicitly not positive. It is important, 
therefore, to generalize the results of~\cite{BDO1} so as to compute the number and properties
of the moduli of vector bundles associated with non-positive spectral covers. 
This will be accomplished in this paper, along with a description of the 
transition moduli.

Specifically, we do the following. In Section 2, the construction of
stable holomorphic vector bundles with structure group $SU(n)$ on simply 
connected elliptic Calabi-Yau threefolds is reviewed. For concreteness, 
we will always work within the context of Calabi-Yau threefolds 
with base surface ${\mathbb F}_{r}$ and $r=0,1,2$.
The expression for the number of 
quark/lepton generations is presented and the condition for anomaly cancellation, 
including the structure of the associated five-branes, is discussed. We close the 
section by giving two explicit examples of anomaly free, $SU(5)$ GUT theories with 
three families of quarks and leptons. 

Instanton moduli for vector bundles constructed 
from spectral data via a Fourier-Mukai transform are discussed in Section 3. 
We begin by giving the definition of these moduli and a generic formula for 
enumerating them. The notion of positive spectral cover is then reviewed and the 
expression for the number of vector bundle moduli in this context, originally 
computed in~\cite{BDO1}, is presented. In the next subsection, we generalize these results 
to arbitrary spectral covers, specifically to the large set of covers that are 
non-positive. Expressions for the number of vector bundle moduli for arbitrary 
spectral covers are comprehensively derived. It is shown
that, unlike the positive case, the spectral 
line bundles may contribute moduli for some non-positive covers. We then 
re-examine the two three family GUT models presented earlier, show that 
they have non-positive spectral covers and compute the number of vector 
bundle moduli in each case. The condition under which a vector bundle 
has the minimal number of moduli for fixed parameters $r$ and $n$ is discussed.
Such a bundle is always non-positive. An explicit formula for the minimal 
number of moduli is presented. 
We close the section by giving an example of a GUT theory with a minimal 
number of moduli, and compare it to the two previous examples.

Section 4 is devoted to the study of transition moduli. We review the 
definition and enumeration of transition moduli in small instanton transitions
involving vector bundles with positive spectral covers. These notions are then 
extended to the case of non-positive spectral covers and a geometric interpretation of 
the associated transition moduli is given, including the role of spectral 
line bundle moduli. Several explicit examples illustrating all of these notions 
are then presented. Concluding remarks are made in Section 5. 
In addition, we give 
important consistency checks of our methods in Appendix A 
and useful linear algebra relations are presented in Appendix B.

An interesting result of our calculations is that vector bundles with 
non-positive spectral covers have far fewer moduli than bundles 
with positive cover, typically by a factor of about a half. Though very promising
in reducing the moduli abundance problem, the number is still rather large. However, 
when our methods are applied to vector bundles over non-simply connected 
Calabi-Yau threefolds,  we expect this number to be dramatically reduced. This will be discussed elsewhere.


\section{Heterotic String on Elliptically Fibered Calabi-Yau Threefolds}


\subsection{Vector Bundles on Elliptically Fibered Calabi-Yau Threefolds}


We consider strongly coupled heterotic string 
theory~\cite{HW1, HW2, Witten96} compactified on
Calabi-Yau threefolds, $X$, structured as elliptic 
curves fibered over a base, B. The maps $\pi:X \rightarrow B$ and $\sigma:B \rightarrow X$ are 
the natural projection and zero section respectively. All such 
Calabi-Yau threefolds are simply connected. The requirement that $X$ be a 
Calabi-Yau space constrains the first Chern class of the tangent bundle to vanish, 
\begin{equation}
c_1(X)=0.
\label{1.3}
\end{equation}
This restricts the base space $B$ to be either del Pezzo, Hirzebruch, Enrique or certain blow-ups of a 
Hirzebruch surface~\cite{MorVafa}. The second Chern class of the tangent 
bundle of $X$ can be shown to be~\cite{FMW}
\begin{equation}
c_2(X) =\pi^* (c_2(B) +11c_1(B)^2) +12 \sigma \cdot \pi^* c_1(B),
\label{1.5}
\end{equation}
where $c_1(B)$ and $c_2(B)$ are the first and second Chern classes of the tangent bundle of $B$.

To specify a heterotic string vacuum, one 
has to present, in addition to the threefold $X$, a 
holomorphic instanton, $A$, satisfying the hermitian Yang-Mills equations,
\begin{eqnarray}
&&F_{ab}=F_{\bar a \bar b}=0, \nonumber \\
&&g^{a \bar b}F_{a \bar b}=0.
\label{1.6}
\end{eqnarray}
If the instanton is indexed in the Lie Algebra of group $G$, then the gauge group 
of the effective four-dimensional theory will be the commutant of $G$ in 
$E_8$. A theorem of Donaldson and Uhlenbeck-Yau~\cite{Donald, Yau} states that the existence of such an instanton is equivalent to constructing a stable, holomorphic vector bundle, $V$, on $X$, in the sense that such a bundle 
admits a connection satisfying~\ref{1.6}.
This theorem also asserts that all possible deformations of the Yang-Mills instanton are 
in one-to-one correspondence with moduli of the stable, holomorphic vector bundle.
Since no hermitian Yang-Mills solutions on Calabi-Yau threefolds have 
been explicitly   constructed,
the vector bundle approach seems to be the only practical way to study the physics 
of instantons and their moduli in heterotic compactifications. A general construction of vector bundles 
on elliptically fibered Calabi-Yau manifolds was given in~\cite{FMW, Donagi}. In this paper, we
consider stable, holomorphic vector bundles with structure group 
$SU(n)$. Such bundles over 
an elliptically fibered Calabi-Yau manifold can be explicitly constructed from two mathematical 
objects, a divisor ${\cal C}$ of $X$, called the spectral cover, and a line bundle 
${\cal N }$ on ${\cal C}$.

A spectral cover is a surface in $X$ that is an $n$-fold cover 
of the base $B$. The general form is 
\begin{equation}
{\cal C}\in |n \sigma +\pi^* \eta|,
\label{1.7}
\end{equation}
where $|n \sigma +\pi^* \eta|$ denotes the linear system of ${\cal O}_{X}(n\sigma + \pi^* \eta)$ and
$\eta$  is a curve class in the base.
For this vector bundle to be stable, we require that the spectral cover be  irreducible.  The linear system
$|n \sigma +\pi^* \eta|$ will contain such irreducible surfaces if the following conditions are met~\cite{FMW, OPP}.
\begin{itemize}
\item
$\pi^{*}\eta$ is an irreducible divisor in $X$. This condition can be satisfied if one demands 
that the linear system $|\eta|$ be base-point free in $B$.
\item
$c_1(\pi^* K_{B}^{\otimes n} \otimes {\cal O}_{X}(\pi^* \eta))$
is effective in $H_4(X, {\mathbb Z})$. This condition is equivalent of saying that 
\begin{equation}
\eta - n c_1(B)
\label{1.7.2}
\end{equation}
is effective in $B$.
\end{itemize}
To make these concepts more concrete, we take the base to be the Hirzebruch surface 
\begin{equation}
B={\mathbb F}_r.
\label{1.8}
\end{equation}
The homology group of ${\mathbb F}_r$ has a basis of effective classes ${\cal S}$ and ${\cal E}$
with intersection numbers
\begin{equation}
{\cal S}^2 =-r, \quad {\cal S} \cdot {\cal E} =1, \quad {\cal E}^2=0.
\label{1.9}
\end{equation}
The first and the second Chern classes of the tangent bundle of ${\mathbb F}_r$
are given by
\begin{equation}
c_1({\mathbb F}_r)=2 {\cal S}+(r+2){\cal E}, \quad
c_2({\mathbb F}_r) =4.
\label{1.9.5}
\end{equation}
Then, in general, ${\cal C}$ is given by~\eqref{1.7} with 
\begin{equation}
\eta= a {\cal S}+b{\cal E},
\label{1.10}
\end{equation}
where $a$ and $b$ are integers. The class $\eta$ is effective if and only if
\begin{equation}
a \geq 0, \quad b\geq 0.
\label{1.11}
\end{equation}
The linear system $|\eta|$ is base point free if and only if~\cite{OPP}
\begin{equation}
b \geq ar.
\label{1.12}
\end{equation}
Finally, $\eta -n c_1(B)$ is effective if and only if
\begin{equation}
a \geq 2n, \quad b\geq n(r+2).
\label{1.13}
\end{equation}
Eqs.~\eqref{1.12} and~\eqref{1.13} guarantee that we can choose ${\cal C}$ to be  an irreducible surface. 
In addition to the spectral cover, it is necessary to specify a line bundle, ${\cal N}$, 
over ${\cal C}$.
In this paper,
we will consider the case when, topologically, such a line 
bundle is a restriction of a global line bundle on $X$ (which we again denote by 
${\cal N}$). For $SU(n)$ bundles, the first Chern class of ${\cal N}$ is given by~\cite{FMW}
\begin{equation}
c_1({\cal N})=n(\frac{1}{2}+\lambda) \sigma +(\frac{1}{2}-\lambda) \pi^* \eta 
+(\frac{1}{2} + n \lambda) \pi^*c_1(B).
\label{1.14}
\end{equation}
Since $c_1({\cal N})$ must be an integer class, it follows that either 
\begin{equation}
\lambda =m+\frac{1}{2} \quad {\rm if} {\ } n {\ } {\rm is {\ } odd}
\label{1.15}
\end{equation}
or
\begin{equation}
\lambda =m, \quad \eta =c_1(B) {\ }{\rm mod}{\ } 2 \quad {\rm if}{\ } n{\ } {\rm is{\ } even},
\label{1.16}
\end{equation}
where $m \in {\mathbb Z}$.

Given these ingredients, the $SU(n)$ vector bundle $V$ can be obtained by the Fourier-Mukai
transform
\begin{equation}
V=\pi_{1*}(\pi^*_2 {\cal N}\otimes {\cal P}).
\label{1.17}
\end{equation}
Here $\pi_1$ and $\pi_2$ are the projections of the fiber product 
$X \times_{B} {\cal C}$ onto the two factors $X$ and ${\cal C}$ 
respectively. ${\cal P}$
is the Poincare line bundle. See~\cite{FMW} for details. The Chern classes 
of $V$ were computed in~\cite{FMW, Curio, Andreas}. The results are
\begin{eqnarray}
&&c_{1}(V)=0, \label{1.18} \\
&&c_2(V)=\sigma \cdot \pi^* \eta -\frac{1}{24} (n^3-n) \pi^* c_1(B) +
\frac{1}{2}(\lambda^2-\frac{1}{4})n\pi^* (\eta \cdot (\eta -n c_1(B))), \label{1.19}\\
&&c_3(V) =2 \lambda \sigma \cdot \pi^* (\eta \cdot (\eta -n c_1(B))).
\label{1.20}
\end{eqnarray}
Finally, we note from the index theorem that
\begin{equation}
N_{gen} =\frac{c_3(V)}{2}
\label{1.21}
\end{equation}
gives the index of the Dirac operator 
in the fundamental representation of $SU(n)$
on a Calabi-Yau threefold and, therefore,
the number of generations in the four-dimensional effective theory.


\subsection{Anomaly Cancellation and Five-Branes}


Any physically acceptable 
heterotic vacuum must be 
anomaly free. In the strongly coupled theory, this condition is given
by~\cite{Witten96}
\begin{equation}
c_2(V_1) +c_2(V_2) +[W]=c_2(X),
\label{1.22}
\end{equation}
where $V_1$ and $V_2$ are the vector bundles on the observable and hidden orbifold fixed planes and
$[W]$ is the class of the holomorphic curve in $X$ around which five-branes 
may be wrapped.
In this paper, for simplicity, we will assume that the vector bundle in the hidden sector, $V_2$, is 
trivial. This simplifies eq.~\eqref{1.22} to 
\begin{equation}
[W]=c_2(X)-c_2(V),
\label{1.23}
\end{equation}
where we denoted $V_1$ by $V$. This equation can be considered as a definition of the 
five-brane class $[W]$. Obviously, the five-brane class must, on physics grounds, be 
represented by an actual curve in $X$. Hence, $[W]$ must be an effective class 
in $H_2(X, {\mathbb Z})$. This condition puts a non-trivial restriction on the 
allowed vector bundle $V$. In the case of elliptically fibered Calabi-Yau threefolds, $[W]$ always 
decomposes as
\begin{equation}
[W]=\sigma \cdot W_B + a_f F,
\label{1,24.5}
\end{equation}
where $W_B=\pi^* w$ is the lift of a curve $w$ in the base,
$a_f$ is an integer and $F$ is the class of the fiber.
$[W]$ will be effective if and only if $w$ is effective 
and $a_f>0$.


\subsection{Examples of Models with Three Generations}


Let us give some examples of the above concepts. We will concentrate 
on GUT models having three generations of quarks and leptons. 

\vspace{0.5cm}
\noindent
{\bf Example 1.} Consider a vector bundle specified by 
$B ={\mathbb F}_1, G=SU(3)$ and the spectral cover 
\begin{equation}
{\cal C} \in | 3\sigma +\pi^*(6 {\cal S} +10{\cal E})|. 
\label{1.25}
\end{equation}
Clearly conditions~\eqref{1.11}-\eqref{1.13} are satisfied.
Let the parameter $\lambda$ of the line bundle ${\cal N}$ be
\begin{equation}
\lambda=\frac{1}{2}.
\label{1.27}
\end{equation}
Using eqs.~\eqref{1.5}, \eqref{1.9}, \eqref{1.9.5}, \eqref{1.19} 
and~\eqref{1.23},
we find that the five-brane class is given by 
\begin{equation}
[W]=\sigma \cdot \pi^* (18 {\cal S} 
+26 {\cal E}) + 96 F.
\label{1.29}
\end{equation}
Since the curve $18{\cal S}+26{\cal E}$ is effective and $96$ is non-negative, 
it follows that  $[W]$ is effective. From~\eqref{1.9}, \eqref{1.20} 
we find 
\begin{equation}
N_{gen}=\frac{1}{2}c_3(V)=3.
\label{1.30}
\end{equation}
{\bf Example 2.} Consider a vector bundle specified by 
$B ={\mathbb F}_1, G=SU(3)$ and the spectral cover 
\begin{equation}
{\cal C} \in | 3\sigma +\pi^* (8 {\cal S} +9{\cal E})|.
\label{1.31}
\end{equation}
Conditions~\eqref{1.11}-\eqref{1.13} are satisfied.
Choose the parameter $\lambda$ of the line bundle ${\cal N}$ be
\begin{equation}
\lambda=\frac{3}{2}.
\label{1.32}
\end{equation}
From eqs.~\eqref{1.5}, \eqref{1.9}, \eqref{1.9.5}, \eqref{1.19} and~\eqref{1.23}
it follows that the five-brane class is given by 
\begin{equation}
[W]=\sigma \cdot \pi^*(16 {\cal S} 
+27 {\cal E}) + 90 F.
\label{1.33}
\end{equation}
Note that $[W]$ is effective. Using~\eqref{1.9}, \eqref{1.20} 
one finds
\begin{equation}
N_{gen}=\frac{1}{2}c_3(V)=3.
\label{1.34}
\end{equation}

We will refer to these examples later in the paper.


\section{Instanton Moduli}


\subsection{General Considerations}


The number of moduli of a stable,
holomorphic $SU(n)$ vector bundle $V$ constructed using the Fourier-Mukai 
transform~\eqref{1.17} is determined by the number of parameters
specifying its spectral cover ${\cal C}$ and by the dimension of the space of holomorphic 
line bundles ${\cal N}$ on ${\cal C}$. In~\cite{BDO1}, it was shown that
the number of parameters of the spectral cover ${\cal C}$ is given by the dimension 
of the linear system $|{\cal C}|$ in $X$. The linear system $|{\cal C}|$
is the projectivization of $H^0(X, {\cal O}_X({\cal C}))$, the space 
of holomorphic sections of the line bundle ${\cal O}_X({\cal C})$. That is 
\begin{equation}
|{\cal C}|={\mathbb P}H^0(X, {\cal O}_X({\cal C})).
\label{2.1}
\end{equation}
Therefore, 
\begin{equation}
dim |{\cal C}|=h^0(X, {\cal O}_X({\cal C})) -1.
\label{2.2}
\end{equation}
This quantity counts the number of parameters specifying the spectral cover ${\cal C}$.
The set of holomorphic line bundles ${\cal N}$ over ${\cal C}$, 
denoted by $Pic({\cal C})$,
is, by definition, determined by the set of holomorphic transition functions
on ${\cal C}$. 
A standard result (see, for example,~\cite{Grif}) says that 
\begin{equation}
dim{\ }Pic({\cal C})=h^{1}({\cal C}, {\cal O}_{{\cal C}}).
\label{2.3}
\end{equation}
Putting this all together, we see that the number of moduli
of a stable, holomorphic $SU(n)$ vector bundle $V$ is given by~\cite{BDO1}
\begin{equation}
n(V)=(h^0(X, {\cal O}_X({\cal C}))-1)+h^{1}({\cal C}, {\cal O}_{{\cal C}}).
\label{2.4}
\end{equation}
Since it will be important later in this paper, let us recall the argument for the identification  \eqref{2.3}. 
It follows from the 
exponential 
sequence on ${\cal C}$ 
\begin{equation}
0 \rightarrow {\mathbb Z} \rightarrow {\cal O}_{{\cal C}} 
\rightarrow {\cal O}_{{\cal C}}^* \rightarrow 0,
\label{2.5}
\end{equation}
where ${\cal O}^*_{{\cal C}}$ is the sheaf of holomorphic nowhere vanishing 
functions,
that
\begin{equation}
0 \rightarrow {H^1(\cal C,{\mathbb Z})} \rightarrow {H^1(\cal C,{\cal O}_{{\cal C}})} 
\rightarrow {Pic^{0}({\cal C})} \rightarrow 0.
\end{equation}
Here $Pic^{0}({\cal C})$ describes the space of all degree zero line bundles on $\cal C$. 
Since $H^1(\cal C,{\mathbb Z})$ is a 
discrete lattice we find $dim{\ }Pic^{0}({\cal C})=h^{1}({\cal C}, {\cal O}_{{\cal C}}) $.
In addition, since $dim{\ }Pic^{0}({\cal C})=dim{\ }Pic({\cal C})$, identification \eqref{2.3} follows.
On a Calabi-Yau threefold $X$, we can describe the vector space $H^1(\cal C,{\cal O}_{{\cal C}})$ rather explicitly. 
Considering the exact sequence 
\begin{equation}
0 \rightarrow {\cal O}_X(-{\cal C}) \rightarrow {\cal O}_{X} 
\rightarrow {\cal O}_{{\cal C}} \rightarrow 0
\label{2.7}
\end{equation}
and the corresponding cohomology sequence
\begin{equation}
\dots \rightarrow H^1(X, {\cal O}_X) 
\rightarrow H^1({\cal C}, {\cal O}_{{\cal C}}) 
\rightarrow H^2(X, {\cal O}_X(-{\cal C}))
\rightarrow H^2(X, {\cal O}_X) \rightarrow \dots,
\label{2.8}
\end{equation}
we obtain
\begin{equation}
H^1({\cal C}, {\cal O}_{{\cal C}}) =H^2(X, {\cal O}_X(-{\cal C})).
\label{2.9}
\end{equation}
Here we have used 
\begin{equation}
H^i(X, \cO_X)=H^{0,i}_{\bar \partial}(X)
\label{2.10}
\end{equation}
together with the fact that on a Calabi-Yau manifold $X$
\begin{equation}
h^{0,1}=h^{0,2}=0.
\label{2.12}
\end{equation}
Furthermore, by Serre duality,
\begin{equation}
H^2(X, {\cal O}_X(-{\cal C}))=
H^1(X, {\cal O}_X({\cal C}))^{*}.
\label{2.13}
\end{equation}
Hence, we find 
\begin{equation}
H^1({\cal C},{\cal O}_{{\cal C}})=H^1(X, {\cal O}_X({\cal C}))^{*}
\label{r1}
\end{equation}
and, in particular, that
\begin{equation}
h^1({\cal C},{\cal O}_{{\cal C}})=h^1(X, {\cal O}_X({\cal C})).
\label{2.13A}
\end{equation}
Therefore, eq.~\eqref{2.4} can be rewritten as 
\begin{equation}
n(V)=(h^0(X, {\cal O}_X({\cal C}))-1)+h^1(X, {\cal O}_X({\cal C})).
\label{2.14}
\end{equation}
This expression suggests that in order to calculate the number of moduli,
one has to learn how to calculate the dimension of the sheaf cohomology groups 
$h^0(X, {\cal O}_X({\cal C}))$ and $h^1(X, {\cal O}_X({\cal C}))$.
In this paper, for specificity,
we will concentrate on Calabi-Yau threefolds elliptically fibered 
over Hirzebruch surfaces ${\mathbb F}_r$ as we did in~\cite{BDO1}. We will also
choose the index $r$ to be $0, 1, 2$. The Calabi-Yau threefolds fibered over
${\mathbb F}_r$ for $r>3$ are singular and require a different approach.
In~\cite{BDO1}, the problem of calculating the dimension of the above sheaf cohomologies 
was simplified
by imposing a special constraint on the spectral cover, called positivity. 
We will review this in the next subsection.
However,
as we will see, most interesting GUT models with three generations 
involve vector bundles whose spectral covers are not positive. 
This is the case, for instance, in both examples in Section 2.  
In this 
paper, we will improve the results of~\cite{BDO1} by giving 
expressions for the number of vector bundle moduli in vacua where the spectral cover is not positive. 


\subsection{Review of Bundles with Positive Spectral Cover}


The simplifying condition on the spectral cover imposed in~\cite{BDO1} was
positivity. By definition, $\cC$ is positive if the first Chern class
of the associated line bundle $\cO_X(\cC)$ can be represented by a positive 
definite two-form. Equivalently, $\cC$ is positive if 
\begin{equation}
\cC \cdot z > 0, 
\label{2.15}
\end{equation}
for every irreducible, effective curve $z$ in $X$. 
In~\cite{BDO1}, it was shown 
that, for Calabi-Yau threefolds elliptically fibered over 
Hirzebruch surfaces, the positivity condition imposes the 
following restrictions on the coefficients $a$ and $b$ in~\eqref{1.7}:
\begin{eqnarray}
&&a>2n, \nonumber \\
&& b>ar+n(2-r). 
\label{2.16}
\end{eqnarray}
Subject to this condition, the calculation of $n(V)$ is greatly simplified. 
It was shown in~\cite{BDO1} that, in this case, 
\begin{equation}
h^i(X, \cO_X(\cC))=0, \quad i>0.
\label{2.17}
\end{equation}
As a consequence, $h^0(X, \cO_X(\cC))$ is equal to the Euler characteristic 
$\chi_E(X, \cO_X(\cC))$ which can be computed using the index theorem
\begin{equation}
\chi_{E}(X, {\cal O}_{X}({\cal C})) =\int_X
ch(\cO_X(\cC)) \wedge Td(X).
\label{2.18}
\end{equation}
The direct evaluation of the right hand side of~\eqref{2.18} gives (see~\cite{BDO1} for details)
\begin{equation}
\chi_E(X, \cO_X(\cC))=\frac{n}{3}(4n^2-1)+nab -(n^2-2)(a+b) +
ar(\frac{n^2}{2}-1) -\frac{n}{2}r a^2
\label{2.18.1}
\end{equation}
and, therefore, 
\begin{equation}
n(V) = \frac{n}{3}(4n^2-1)+nab -(n^2-2)(a+b) +
ar(\frac{n^2}{2}-1) -\frac{n}{2}r a^2 -1.
\label{2.19}
\end{equation}
The expression~\eqref{2.18.1} for $\chi_E(X, \cO_X(\cC))$ is valid for all choices of $a$ and $b$, whereas
the expression for $n(V)$~\eqref{2.19} is valid only if inequalities~\eqref{2.16} are satisfied. 
However, it is straightforward to check that in 
the three families examples presented in Section 2, the integers $a$ 
and $b$ do not 
satisfy eq.~\eqref{2.16}. Thus, unfortunately, many physically 
interesting vacua do not arise from 
bundles with positive spectral cover. 


\subsection{Bundles with Arbitrary Irreducible Spectral Cover}


In this subsection,the positivity condition will not be imposed. Therefore, 
we have to find an alternative procedure for calculating 
$h^0(X, \cO_X(\cC))$ and $h^1(X, \cO_X(\cC))$. For specificity, we consider 
Calabi-Yau threefolds elliptically fibered over ${\mathbb F}_r$. Let us recall the 
structure of  ${\mathbb F}_r$. The Hirzebruch surface ${\mathbb F}_r$ is 
a ${\mathbb P}^1$ bundle over ${\mathbb P}^1$, ${\mathbb F}_0$
being just the trivial product ${\mathbb P}^1 \times {\mathbb P}^1$. 
The base of this bundle is the curve ${\cal S} \simeq {\mathbb P}^1$ and the 
fiber is the curve ${\cal E} \simeq {\mathbb P}^1$. Their intersection numbers are 
given in eq.~\eqref{1.9}. Any spectral cover will be of the form 
\begin{equation}
\cC \in |n \sigma +\pi^* \eta|,
\label{2.20.1}
\end{equation}
where
\begin{equation}
\eta= a {\cal S} +b {\cal E}
\label{2.20.1a}
\end{equation}
and the coefficients $a$ and $b$ satisfy eqs.~\eqref{1.11}-\eqref{1.13}. 
Recall that $n$ is the rank of the vector bundle.
Let us denote the projection of ${\mathbb F}_r$ onto the base ${\cal S}$ by 
$\rho$
\begin{equation}
\rho : B={\mathbb F}_r \rightarrow {\cal S}.
\label{2.21}
\end{equation}
The natural idea will be to relate $h^0(X, \cO_X(\cC))$ and $h^1(X, \cO_X(\cC))$ to 
cohomology groups on ${\cal S} \simeq {\mathbb P}^1$ and then use the standard expressions
\begin{equation}
h^0({\mathbb P}^1, \cO_{{\mathbb P}^1}(m)) =m+1, \quad m \geq 0
\label{2.22}
\end{equation}
and 
\begin{equation}
h^0({\mathbb P}^1, \cO_{{\mathbb P}^1}(m)) =0, \quad m \leq -1
\label{2.22.1}
\end{equation}
as well as
\begin{equation}
h^1({\mathbb P}^1, \cO_{{\mathbb P}^1}(m)) =0, \quad m \geq 0
\label{2.23.1}
\end{equation}
and
\begin{equation}
h^1({\mathbb P}^1, \cO_{{\mathbb P}^1}(m)) =-m-1, \quad m \leq -1.
\label{2.23}
\end{equation}
By definition
\begin {equation}
H^0 (X, \cO_X(\cC))=H^0 (B, \pi_*\cO_X(\cC)).
\label{2.24}
\end{equation}
Using a Leray spectral sequence, we find 
\begin {equation}
H^1 (X, \cO_X(\cC))=H^1 (B, \pi_*\cO_X(\cC)). 
\label{2.25}
\end{equation}
To justify~\eqref{2.25}, we have to show that
\begin{equation}
H^0(B, R^1\pi_*  \cO_X(\cC)) =0.
\label{2.26}
\end{equation}
This can be seen from the following simple calculation.
By definition, for every point $p$ in the base $B={\mathbb F}_r$ we have 
\begin{equation}
R^1\pi_* \cO_X(\cC)|_{p\in B} =H^1(F_p, \cO_X(\cC)|_{F_p}),
\label{2.27}
\end{equation}
where $F_p$ is the elliptic fiber over $p$. For a spectral cover of the 
form~\eqref{2.20.1}, \eqref{2.20.1a}
we have 
\begin{equation}
\cO_{F_p}(\cC|_{F_p}) =\cO_{F_p} ((n\sigma +\pi^* (a {\cal S} + b {\cal E}))\cdot F)=
\cO_{F_p}( n \sigma(p)).
\label{2.28}
\end{equation}
Here, we have used the intersection properties
\begin{equation}
\s \cdot F=\sigma(p), \quad \pi^*{\cal S} \cdot F=   \pi^*{\cal E} \cdot F=0.
\label{2.29}
\end{equation}
On any irreducible curve, $H^1$ with coefficients in any line bundle of positive degree vanishes
\begin{equation}
H^1(F_p, \cO_{F_p}( n \sigma(p)))=0.
\label{2.30}
\end{equation}
We will assume in this paper that all fibers of $\pi$ are irreducible.
Then, from eq.~\eqref{2.27} we conclude that the sheaf 
$R^1\pi_* \cO_X(\cC)$ is the zero sheaf,
\begin{equation}
R^1\pi_* \cO_X(\cC) =0,
\label{zero}
\end{equation}
since it vanishes at every point $p \in B$.
This proves eq.~\eqref{2.26} and, hence, eq.~\eqref{2.25}. 

To proceed, we need to calculate
$\pi_* \cO_X(\cC)$. Since
\begin{equation}
\pi_* \cO_X(\cC) = \pi_* \cO_X(n\sigma +\pi^* \eta) =
\pi_* \cO_X(n \sigma) \otimes \cO_B (\eta),
\label{2.31}
\end{equation}
it suffices to calculate $\pi_* \cO_X(n \sigma)$. This can be done by induction.
Take $n=1$ and consider the following sequence
\begin{equation}
0\rightarrow \cO_X \rightarrow 
\cO_X(\sigma)\rightarrow  \cO_X(\sigma)|_{\sigma} \rightarrow 0.
\label{2.32}
\end{equation}
Inclusion $\cO_X \hookrightarrow \cO_X(\s)$ induces the sheaf map
\begin{equation}
i:\pi_* \cO_X \rightarrow \pi_* \cO_X (\s).
\label{2.34}
\end{equation}
Restricted to any point $p\in B$, this map becomes 
\begin{equation} 
i_p: H^0(F_p, \cO_{F_p}) \rightarrow H^0(F_p, \cO_{F_p}(\s)|_{F_p}),
\label{2.35}
\end{equation}
where we have used the fact that $\pi_{*}{\cal{O}}_{X}= {\cal{O}}_{B}$.
Since $\s \cdot F=1$, the degree of $\cO_{F_p}(\s)|_{F_p}$ is unity and, 
hence, it has 
a unique holomorphic section. Then the map $i_p$ is an isomorphism. 
Since this is true at any point $p \in B$, it follows that
\begin{equation}
\pi_* \cO_X(\s) =\pi_* \cO_X =\cO_B.
\label{2.36}
\end{equation}
Now take $n=2$ and consider the sequence
\begin{equation}
0\rightarrow \cO_X (\s)\rightarrow 
\cO_X(2\sigma)  \rightarrow  \cO_X(2\sigma)|_{\sigma} \rightarrow 0.
\label{2.37}
\end{equation}
Using~\eqref{2.36}, the relation
\begin{equation}
\s \cdot \s =-\pi^*c_1(B) \cdot \s
\label{2.38}
\end{equation}
and the fact that
\begin{equation}
R^1 \pi_*\cO_X (\s) =0,
\label{2.40}
\end{equation}
the direct image of the sequence~\eqref{2.37} can be expressed as 
\begin{equation}
0\rightarrow \cO_B \rightarrow 
\pi_* \cO_X(2\sigma) \rightarrow \cO_B(-2 c_1 (B)) \rightarrow 0.
\label{2.39}
\end{equation}
Relation~\eqref{2.38} was proven in~\cite{FMW}, whereas~\eqref{2.40} follows 
as a special case of eq.~\eqref{2.26} with $a=b=0$. From eq.~\eqref{2.39},
we would like to conclude that 
\begin{equation}
\pi_* \cO_X (2 \s) =\cO_B \oplus \cO_B (-2 c_1(B)).
\label{2.41}
\end{equation}
This is indeed the case provided the extension group
\begin{equation}
Ext_B^1(\cO_B(-2c_1(B)), \cO_B) \simeq H^1(B, \cO_B(2c_1(B)))=0.
\label{2.42}
\end{equation}
We will prove this below for the relevant case of $B={\mathbb F}_r, r=0, 1, 2$.
Continuing in this way, we obtain
\begin{equation}
\pi_* \cO_X (n \s) =\cO_B \oplus 
{\bigoplus_{i=2}^{n}} \cO_B (-i c_1 (B)).
\label{2.43}
\end{equation}
It follows from this and~\eqref{2.31} that
\begin{equation}
\pi_* {\cO_X} ({\cal{C}}) = {\cO_B} (\eta){ \oplus}  \bigoplus_{i=2}^{n} \cO_B (\eta -i c_1 (B)).
\label{2.44}
\end{equation}
In the previous section, we stated that the stability of the bundle 
demands that
$\eta -n c_1 (B)$ be effective. It follows that every line bundle on 
the right hand side of
eq.~\eqref{2.44} must be of the form 
\begin{equation}
\cO_B(c{\cal S} +d{\cal E}), \quad c \geq 0, \quad d \geq 0.
\label{2.45}
\end{equation}
We now use the second projection $\rho$ given by eq.~\eqref{2.21} to relate 
data on $B={\mathbb F}_r$ to data on ${\cal S} \simeq {\mathbb P}_1$. 
Note that by definition it follows that
\begin{equation}
H^0 (B, \cO_B(c{\cal S} +d{\cal E})) = H^0 ({\cal S}, \rho_* \cO_B(c{\cal S} +d{\cal E}))
\label{2.46}
\end{equation}
and from a Leray spectral sequence argument, that
\begin{equation}
H^1 (B, \cO_B(c{\cal S} +d{\cal E})) = H^1 ({\cal S}, \rho_* \cO_B(c{\cal S} +d{\cal E})).
\label{2.47}
\end{equation}
In eq.~\eqref{2.47}, we have used the fact that 
\begin{equation}
R^1 \rho_* \cO_B (c {\cal S} +d{\cal E}) =0, \quad c, d \geq 0.
\label{2.48}
\end{equation}
This can be proven is a similar way to eq.~\eqref{zero}.
Combining these results with~\eqref{2.24}, \eqref{2.25}, \eqref{2.44} 
and~\eqref{2.45}, we have 
\begin{equation}
H^0 (X, \cO_X(\cC))=H^0({\cal S}, \rho_* \pi_*  \cO_X(\cC))
\label{2.48.1}
\end{equation}
and 
\begin{equation}
H^1 (X, \cO_X(\cC))=H^1({\cal S}, \rho_* \pi_*  \cO_X(\cC)).
\label{2.48.2}
\end{equation}
Therefore, by computing $\rho_* \cO_B (c {\cal S} + d{\cal E})$
for $c,d \geq 0$, one can relate 
the cohomology groups 
$H^0(X, \cO_X(\cC))$ and $H^1(X, \cO_X(\cC))$ to cohomology groups on the curve ${\cal S}$ 
by means of eqs.~\eqref{2.48.1} and~\eqref{2.48.2}.
We begin by noting that
\begin{equation}
\rho_* \cO_B(c{\cal S} + d{\cal E}) =
\rho_* \cO_B(c {\cal S}) \otimes \cO_{{\cal S}}(d).
\label{2.49}
\end{equation}
Hence, it suffices to compute $\rho_* \cO_B (c \cS)$ for $c\geq 0$.
This can be accomplished by an induction procedure similar to that used for
$\pi_* \cO_X (n \s)$.
Obviously, for $c=0$ 
\begin{equation}
\rho_* \cO_B =\cO_{\cS}.
\label{2.49.5}
\end{equation}
Take $c=1$ and consider the short exact sequence 
\begin{equation}
0 \rightarrow \cO_B  \rightarrow \cO_B(\cS) \rightarrow \cO_B(\cS)|_{\cS} 
\rightarrow 0.
\label{2.50}
\end{equation}
Taking its direct image, we get 
\begin{equation}
0 \rightarrow \cO_{\cS} \rightarrow \rho_* \cO_B({\cS})  
\rightarrow \cO_{\cS}(-r) \rightarrow 0,
\label{2.51}
\end{equation}
where $r$ is the subscript of ${\mathbb F}_r$ and we have used eqs.~\eqref{1.9} and~\eqref{2.48}.
One can easily show that 
\begin{equation}
Ext^1_{{\mathbb P}^1}(\cO_{\cS}(-r),\cO_{\cS})=H^1({\mathbb P}^1, \cO_{\cS}(r))=0
\end{equation}
for positive $r$
Hence, we find
\begin{equation}
\rho_* \cO_B({\cS}) = \cO_{\cS} \oplus \cO_{\cS} (-r).
\label{2.52}
\end{equation}
Now take $c=2$ and consider the sequence 
\begin{equation}
0 \rightarrow \cO_B(\cS)  \rightarrow \cO_B(2\cS) \rightarrow \cO_B(2\cS)|_{\cS} 
\rightarrow 0.
\label{2.53}
\end{equation}
Taking the direct image, and showing that the corresponding extension space vanishes, we get
\begin{equation}
0 \rightarrow \cO_{\cS} \oplus \cO_{\cS} (-r)  \rightarrow \rho_* \cO_B(2{\cS})  
\rightarrow \cO_{\cS}(-2r) \rightarrow 0,
\label{2.54}
\end{equation}
where eqs.~\eqref{2.48} and~\eqref{2.52} have been used. 
Then
\begin{equation}
\rho_* \cO_B({2\cS}) = \cO_{\cS} \oplus \cO_{\cS} (-r) \oplus \cO_{\cS} (-2r).
\label{2.55}
\end{equation}
Continuing in this way, we obtain
\begin{equation}
\rho_* \cO_B({c\cS}) = \cO_{\cS} \oplus \bigoplus_{j=1}^c \cO_{\cS}(-jr).
\label{2.56}
\end{equation}
It follows from this and~\eqref{2.49} that
\begin{equation}
\rho_* \cO_B({c\cS}+d\cE) = \cO_{\cS}(d) \oplus \bigoplus_{j=1}^c \cO_{\cS}(d-jr).
\label{2.57}
\end{equation}
Eqs.~\eqref{2.44}, \eqref{2.48.1}, \eqref{2.48.2} and~\eqref{2.57}
provide a complete relation between data on $X$ and data on ${\mathbb P}^1$. 
We will use them to calculate $h^0(X, \cO_X(\cC))$ and $h^1(X, \cO_X(\cC))$.

First, however, we will prove~\eqref{2.42}.
From~\eqref{1.9.5}, 
\begin{equation}
H^1(B, \cO_B(2 c_1 (B))) = H^1(B, \cO_B (4 \cS +2(r+2)\cE)).
\label{2.58}
\end{equation}
Using eqs.~\eqref{2.47} and~\eqref{2.57}, it follows that
\begin{eqnarray}
&&H^1(B, \cO_B (4 \cS +2(r+2)\cE))= H^1(\cS, \rho_*\cO_B (4 \cS +2(r+2)\cE)) = \nonumber \\
&&H^1(\cS, \cO(2(r+2)) \oplus 
\bigoplus_{j=1}^4 \cO_{\cS}(2(r+2) -jr)).
\label{2.59}
\end{eqnarray}
Note that in the last expression each line bundle on ${\cal S}$ 
has non-negative degree for $r=0, 1, 2$. Therefore, by eqs.~\eqref{2.23} 
and~\eqref{2.23.1}, 
each has vanishing $H^1$. This proves~\eqref{2.42}.

We now have all ingredients necessary  to compute the number of moduli 
for a vector bundle specified by a spectral cover 
\begin{equation}
{\cal C}\in |n \s +\pi^*(a \cS +b\cE)|.
\label{2.60}
\end{equation}
Recall, that the conditions on the coefficients $a$ and $b$ are 
\begin{equation}
a, b \geq 0, \quad b \geq ar, \quad a \geq 2n, \quad b \geq n(r+2).
\label{2.61}
\end{equation}
We need to compute the form of the line bundles that arise in
$\rho_* \pi_* \cO_X(\cC)$. From eqs.~\eqref{2.44}, \eqref{1.9.5} 
and~\eqref{2.60} we have 
\begin{equation}
\pi_* \cO_X(\cC) =\cO_B(a \cS+b\cE) \oplus
\bigoplus_{i=2}^n \cO_B ((a-2i)\cS +(b-(r+2)i)\cE).
\label{2.64}
\end{equation}
Now consider the $\rho$-direct image of all the line bundles on the 
right hand side of~\eqref{2.64}. 
Using~\eqref{2.57}, we see that
\begin{equation}
\rho_* \cO_B(a \cS+b\cE) =\cO_{\cS} (b) \oplus \bigoplus_{j=1}^a \cO_{\cS} (b-jr).
\label{2.65}
\end{equation}
Note from~\eqref{2.61} that all degrees on the right hand side of~\eqref{2.65} are non-negative.
Similarly, ~\eqref{2.57} implies
\begin{equation}
\rho_* \cO_B((a-2i) \cS+(b-(r+2)i\cE) =\cO_{\cS} (b-(r+2)i) \oplus \bigoplus_{j=1}^{a-2i} \cO_{\cS} (b-(r+2)i-jr).
\label{2.66}
\end{equation}
It is easy to see that if $r=0, 2$, all degrees on the 
right hand side of eq.~\eqref{2.66} are non-negative as a consequence of eq.~\eqref{2.61}.
Indeed, if $r=0$, all degrees in eq.~\eqref{2.66} are $b-2i$. Since $i$ runs from $2$ to $n$, the 
minimal possible degree is $b-2n$. But from eq.~\eqref{2.61} we see that $b$ is always greater or equal $2n$.
Similarly, when $r=2$, the minimal possible degree in~\eqref{2.66} is when $j$ is maximal, that is, 
$b-4i-2(a-2i)=b-2a$. But from eq.~\eqref{2.61} we see that $b-2a$ is always non-negative for $r=2$.
On the contrary, for $r=1$, $\rho_* \pi_* \cO_X(\cC)$ contains line bundles 
of both positive and negative degree. We find it convenient, therefore, to separate the cases $r=0, 2$ and $r=1$.
In the first case, the calculation of $h^0(X, \cO_X(\cC))$ and $h^1(X, \cO_X(\cC))$ will be 
relatively easy, whereas in the latter case it will require greater care.\\ 

{\noindent\bf $r=0, 2$ case:}\\

Let us begin with 
the case $r=0, 2$. We have just concluded that, in this case, $\rho_* \pi_* \cO_X(\cC)$
contains only line bundles on $\cS$ of non-negative degree. Then from eqs.~\eqref{2.23} and~\eqref{2.23.1}
it follows that
\begin{equation}
h^1(X, \cO_X(\cC)) =0.
\label{2.67}
\end{equation}
That is, as in the case of bundles with positive spectral cover, the line bundles $\cN$ do not 
have any continuous parameters. Now calculate $h^0(X, \cO_X(\cC))$. Using
~\eqref{1.9.5},~\eqref{2.24}, and~\eqref{2.44}, we have 
\begin{equation}
h^0(X, \cO_X(\cC))=
h^0(B, \cO_B(a \cS+b\cE))+ 
\sum_{i=2}^n h^{0}(B,\cO_B ((a-2i)\cS +(b-(r+2)i)\cE)).
\label{2.68}
\end{equation}
Consider the first term. By eq.~\eqref{2.65}
\begin{equation}
h^0(B, \cO_B(a \cS+b\cE))=h^0(\cS, \cO_{\cS}(b)) +
\sum_{j=1}^{a}h^0(\cS, \cO_{\cS}(b-rj)).
\label{2.69}
\end{equation}
Since all degree in eq.~\eqref{2.69} are positive, it follows from from~\eqref{2.22} that
\begin{equation}
h^0(B, \cO_B(a \cS+b\cE)) =(b+1) +\sum_{j=1}^{a}(b-rj+1)= (a+1)(b+1-\frac{ra}{2}).
\label{2.70}
\end{equation}
Similarly,
\begin{eqnarray} 
&&\sum_{i=2}^n h^{0}(B,\cO_B ((a-2i)\cS +(b-(r+2)i)\cE))=
\sum_{i=2}^n h^0(\cS, \cO_{\cS}(b-(r+2)i)) +
\nonumber \\
&&
\sum_{i=2}^n
h^0(\cS, \bigoplus_{j=1}^{a-2i} \cO_{\cS} (b-(r+2)i-jr)).
\label{2.70.1}
\end{eqnarray}
Since all degrees here are non-negative, 
we find using eq.~\eqref{2.22} that
\begin{eqnarray}
&&\sum_{i=2}^n h^{0}(B,\cO_B ((a-2i)\cS +(b-(r+2)i)\cE))= \nonumber \\
&&\sum_{i=2}^n(b-(r+2)i+1) +  
\sum_{i=2}^n\sum_{j=1}^{a-2i}(b-(r+2)i-jr+1)= 
\nonumber \\
&&
\frac{1}{2}(n-1)(nr-2n-2r+2b-2)
+\nonumber \\
&&
\frac{1}{6}(n-1)(12+ 6a + 12b - 6ab - 14n + 6an + 6bn - 8 n^2 - 
6r - 3ar + \nonumber \\
&&3a^2r - 3nr - 3anr).
\label{2.71}
\end{eqnarray}
Adding eqs.~\eqref{2.70} and~\eqref{2.71}, we get
\begin{equation}
h^0(X, \cO_X(\cC))=
\frac{n}{3}(4n^2-1)+nab -(n^2-2)(a+b) +
ar(\frac{n^2}{2}-1) -\frac{n}{2}r a^2.
\label{2.72}
\end{equation}
Combining~\eqref{2.14},~\eqref{2.67} and ~\eqref{2.72}, 
the number of moduli is given by 
\begin{equation}
n(V)=\frac{n}{3}(4n^2-1)+nab -(n^2-2)(a+b) +
ar(\frac{n^2}{2}-1) -\frac{n}{2}r a^2-1.
\label{2.73}
\end{equation}
This is the final expression for the number of instanton moduli for 
$r=0, 2$. Surprisingly, it coincides with eq.~\eqref{2.19} obtained 
for bundles with positive spectral cover. In addition, eq.~\eqref{2.72}
coincides with eq.~\eqref{2.18.1} for the Euler characteristic. That is
\begin{equation}
h^0(X, \cO_X(\cC))=\chi_E(X, \cO_X(\cC)).
\label{2.74}
\end{equation}
It follows that not only does $h^1(X, \cO_X(\cC))$ vanish, as we 
have shown, but 
\begin{equation}
h^2(X, \cO_X(\cC))=h^3(X, \cO_X(\cC))=0
\label{2.74.1}
\end{equation}
must be true as well. 
As a consistency check of our results, 
we will prove in Appendix A, using independent methods, that this is indeed the case.\\ 
 
{\noindent \bf $r=1$ case:}\\

Let us now consider the more complicated case when $r=1$.
In this case, $h^1(X, \cO_X(\cC))$ will generically be non-zero and line 
bundles on $\cC$ will contribute instanton moduli. 
To begin with, we consider eq.~\eqref{2.66}.
This will 
contain only line bundles of non-negative degree 
if (see eq.~\eqref{2.23.1})
\begin{equation}
b-3i-j \geq -1
\label{2.0.67}
\end{equation}
for all allowed values of $i$ and $j$. Inequality~\eqref{2.0.67}
is satisfied if and only if
\begin{equation}
b \geq a+n-1.
\label{2.0.68}
\end{equation}

{\noindent \bf i) $b \geq a+n -1$:}\\

Assume that $a$,$b$ satisfy~\eqref{2.0.68}. Then it follows 
from~\eqref{2.23.1} that 
$h^{1}(X,{\cal{O}}_{X}(\cal{C}))$ will not contribute to $n(V)$.
In this case, the entire contribution to $n(V)$ comes from $h^0(X, \cO_X(\cC))$ as before.
A calculation identical to that done in the case $r=0, 2$ again leads eq.~\eqref{2.73}.
We conclude that if
\begin{equation}
r=1, \quad b \geq a+n-1,
\label{2.0.69}
\end{equation}
the number of moduli is given by 
\begin{equation}
n(V)=\frac{n}{3}(4n^2-1)+nab -(n^2-2)(a+b) +
ar(\frac{n^2}{2}-1) -\frac{n}{2}r a^2-1.
\label{2.0.70}
\end{equation}
As before, $h^0(X, \cO_X(\cC))$ coincides with $\chi_E$.\\

{\noindent\bf ii) $b< a+n-1$:}\\ 

Now assume that
\begin{equation}
b<a+n-1.
\label{2.0.71}
\end{equation}
Let us first calculate $h^1(X, \cO_X(\cC))$. From eq.~\eqref{2.68}, 
$\pi_* \cO_X (\cC)$ contains two contributions
\begin{equation}
\cO_B(a \cS+b\cE)
\label{2.0.72}
\end{equation}
and 
\begin{equation}
\bigoplus_{i=2}^n \cO_B ((a-2i)\cS +(b-(r+2)i)\cE)).
\label{2.0.73}
\end{equation}
Using~\eqref{2.65}, the direct image with respect to $\rho$ 
of~\eqref{2.0.72} is 
\begin{equation}
\rho_*\cO_B(a \cS+b\cE) =\cO_{\cS}(b) \oplus \bigoplus_{j=1}^{a}\cO_{\cS}(b-j)
\label{2.0.74}
\end{equation}
which does not contain line bundles of degree less than $-1$. Hence, its contribution 
to $h^1(X, \cO_X(\cC))$ vanishes. Now consider the $\rho$-direct 
image of line bundles in~\eqref{2.0.73}. From eq.~\eqref{2.66} we have
\begin{eqnarray}
&&\rho_* \cO_B((a-2i) \cS+(b-3i\cE) =\cO_{\cS} (b-3i) \oplus \bigoplus_{j=1}^{a-2i} \cO_{\cS} (b-3i-jr)=
\nonumber \\
&&\cO_{\cS} (b-3i) \oplus \cO_{\cS} (b-3i-1) \oplus \dots \oplus \cO_{\cS} (b-a-1).
\label{2.76}
\end{eqnarray}
In order for $h^1(\cS, \rho_* \cO_B((a-2i) \cS+(b-3i\cE)))$ to be non-zero, at least one 
degree in eq.~\eqref{2.76} must be less than $-1$. For $i \leq b-a+1$, all degrees in
eq.~\eqref{2.76} are greater than $-1$ and 
\begin{equation}
h^1(\cS, \rho_* \cO_B((a-2i) \cS+(b-3i\cE)))=0.
\label{2.77}
\end{equation}
For $i=b-a+2$, we have 
\begin{equation}
\rho_* \cO_B((a-2i) \cS+(b-3i\cE)) =\cO_{\cS}(-2) \oplus  \cO_{\cS}(-1) \oplus \cO_{\cS} \oplus \dots,
\label{2.78}
\end{equation}
where the ellipsis denote the line bundles 
of positive degree. Therefore, by eqs.~\eqref{2.23} and~\eqref{2.23.1}
we get 
\begin{equation}
h^1(\cS, \rho_* \cO_B((a-2i) \cS+(b-3i\cE)))=1.
\label{2.79}
\end{equation}
More generally, for $b-a+1 <i \leq n$ we have 
\begin{equation}
\rho_* \cO_B((a-2i) \cS+(b-3i\cE)) =\cO_{\cS}(b-a-i) \oplus
\cO_{\cS}(b-a+1-i) \oplus \cO_{\cS}(-1) \oplus \cO_{\cS} \oplus \dots,
\label{2.80}
\end{equation}
where the ellipsis denote line bundles of positive degree. 
From eqs.~\eqref{2.80}, \eqref{2.23} and~\eqref{2.23.1} we find
\begin{equation}
h^1(\cS, \rho_* \cO_B((a-2i) \cS+(b-3i\cE)))=
\sum_{j=2}^{i-b+a}(j-1)=\frac{(i-b+a)(i-b+a-1)}{2}.
\label{2.81}
\end{equation}
To find $h^1(X, \cO_X(\cC))$, we have to sum over $i$ from $b-a+2$ to $n$. The result is
\begin{eqnarray}
&&h^1(X, \cO_X(\cC)) =\sum_{i=b-a+2}^{n}\frac{(i-b+a)(i-b+a-1)}{2}= \nonumber \\
&&\frac{1}{6}(n-b+a)(n-b+a-1)(n-b+a+1).
\label{2.82}
\end{eqnarray}
This concludes our calculation of $h^1(X, \cO_X(\cC))$.
Similarly, we can compute $h^0(X, \cO_X(\cC))$.
It follows from eqs.~\eqref{2.22} and~\eqref{2.22.1} that
in the $\rho$-direct image of eqs.~\eqref{2.0.72} and~\eqref{2.0.73}
only line bundles of non-negative degree will contribute.
First, 
\begin{equation}
h^0(B, \cO_B(a \cS+b\cE)) =(a+1)(b+1-\frac{ra}{2}),
\label{2.83}
\end{equation}
as was computed previously in~\eqref{2.70}.
Furthermore, as was discussed earlier, for $i \leq b-a$
the degrees of all line bundles in~\eqref{2.76} are positive. Therefore,
using eqs.~\eqref{2.76}, \eqref{2.22} and~\eqref{2.22.1},
for $i \leq b-a$ we have
\begin{eqnarray}
&&h^0(\cS, \rho_* \cO_B((a-2i) \cS+(b-3i\cE)))=
\sum_{i=2}^{b-a} \sum_{j=b-a-i}^{b-3i}(j+1) =\nonumber \\
&&\frac{1}{6}(b-a-1)(6-11a+8a^2+2b-7ab+2b^2).
\label{2.84}
\end{eqnarray}
Now consider the case $i \geq b-a+1$. Then, it follows from~eq.~\eqref{2.76} that
\begin{equation}
\rho_* \cO_B((a-2i) \cS+(b-3i\cE)) ={\cal L}_+\oplus {\cal L}_{-},
\label{2.85}
\end{equation}
where ${\cal L}_{+}$ is the direct sum of the line bundle of non-negative degree
\begin{equation}
{\cal L}_{+}=\cO_{\cS}(b-3i) \oplus \dots \oplus \cO_{\cS}(1) \oplus \cO_{\cS}
\label{2.86}
\end{equation}
and ${\cal L}_{-}$ is the direct sum of line bundles of negative degree.
From eqs.~\eqref{2.22} and~\eqref{2.22.1} we find
\begin{eqnarray}
&&h^0(\cS, \rho_* \cO_B((a-2i) \cS+(b-3i)\cE))=h^0(\cS, {\cal L}_{+})=
\sum_{i=b-a+1}^{n}\sum_{j=0}^{b-3i} (j+1) =\nonumber \\
&&\frac{1}{2} (n-b+a)(3a^2+b^2 -1 -3ab-3an+3n^2).
\label{2.87}
\end{eqnarray}
Now combine the results of eqs.~\eqref{2.83}, \eqref{2.84} and~\eqref{2.87}
to obtain
\begin{eqnarray}
&&h^0(X, \cO_X(\cC))=\frac{n}{3}(4n^2-1)+nab -(n^2-2)(a+b) +
ar(\frac{n^2}{2}-1) -\frac{n}{2}r a^2 + \nonumber \\
&& \frac{1}{6}(n-b+a)(n-b+a-1)(n-b+a+1).
\label{2.88}
\end{eqnarray}
Using eqs.~\eqref{2.14},~\eqref{2.82} and~\eqref{2.88},
we find that the number of moduli in the case 
\begin{equation}
r=1, \quad b<a+n-1
\label{2.88.2}
\end{equation}
is given by 
\begin{eqnarray}
&&n(V)=\frac{n}{3}(4n^2-1)+nab -(n^2-2)(a+b) +
ar(\frac{n^2}{2}-1) -\frac{n}{2}r a^2 -1+ \nonumber \\
&& \frac{1}{3}(n-b+a)(n-b+a-1)(n-b+a+1).
\label{2.89}
\end{eqnarray}
Note that, using eqs.~\eqref{2.18.1} and~\eqref{2.82}, the last expression can be rewritten as
\begin{equation}
h^0(X, \cO_X(\cC)) =\chi_E(X, \cO_X(\cC)) +h^1(X, \cO_X(\cC)).
\label{2.88.1}
\end{equation}
This implies that both $h^2(X, \cO_X(\cC))$ and $h^3(X, \cO_X(\cC))$ must 
vanish. We will check our results by presenting
an independent proof of this in Appendix A.. 
Before concluding this subsection, let us summarize the results. First,
${\cal{C}}$ being effective and irreducible requires that coefficients $a$,$b$ 
satisfy~\eqref{2.61} for any choice of $n$ and $r$.
\begin{itemize}
\item 
If
\begin{equation}
r=0, 2 \quad {\rm or} \quad r=1, b \geq a+n-1
\label{2.89.1}
\end{equation}
then 
\begin{equation}
h^0(X, \cO_X(\cC))=
\frac{n}{3}(4n^2-1)+nab -(n^2-2)(a+b) +
ar(\frac{n^2}{2}-1) -\frac{n}{2}r a^2, 
\label{1}
\end{equation}
\begin{equation}
h^1(X, \cO_X(\cC))=0, 
\label{2}
\end{equation}
and
\begin{equation}
n(V)= \frac{n}{3}(4n^2-1)+nab -(n^2-2)(a+b) +
ar(\frac{n^2}{2}-1) -\frac{n}{2}r a^2-1.
\label{2.90}
\end{equation}

\item 
If
\begin{equation}
r=1, \quad b < a+n-1
\label{2.90.1}
\end{equation}
then 
\begin{eqnarray}
&&h^0(X, \cO_X(\cC))=
\frac{n}{3}(4n^2-1)+nab -(n^2-2)(a+b) +
ar(\frac{n^2}{2}-1) -\frac{n}{2}r a^2 + \nonumber \\
&&\frac{1}{6}(n-b+a)(n-b+a-1)(n-b+a+1),
\label{3}
\end{eqnarray}
\begin{equation}
h^1(X, \cO_X(\cC))=\frac{1}{6}(n-b+a)(n-b+a-1)(n-b+a+1), 
\label{4}
\end{equation}
and
\begin{eqnarray}
&&n(V)= \frac{n}{3}(4n^2-1)+nab -(n^2-2)(a+b) +
ar(\frac{n^2}{2}-1) -\frac{n}{2}r a^2-1+\nonumber \\
&&\frac{1}{3}(n-b+a)(n-b+a-1)(n-b+a+1).
\label{2.91}
\end{eqnarray}
\end{itemize}
Eqs.~\eqref{2.90}, \eqref{2.91} give a complete classification of the number 
of instanton moduli on Calabi-Yau manifolds elliptically fibered over the 
Hirzebruch surfaces ${\mathbb F}_r$ with $r=0,1,2$.


\subsection{Examples with Non-Positive Spectral Covers}


Let us give some examples of the above concepts. We will consider again the two GUT 
models with three generations presented as {\bf Example 1} and {\bf Example 2} in 
Section 2. 

\vspace{0.5cm}
\noindent
{\bf Example 1.}
Consider the vector bundle specified by $B={\mathbb F}_1, G=SU(3)$, the 
spectral cover
\begin{equation}
{\cal C}\in |3 \s +\pi^*(6\cS+10\cE)|
\label{2.116}
\end{equation}
and 
\begin{equation}
\lambda=\frac{1}{2}.
\label{2.117}
\end{equation}
As discussed previously, coefficients $a=6, b=10$ do not satisfy 
condition~\eqref{2.16} and, hence, spectral cover~\eqref{2.116} 
is not positive.
Using eqs.~\eqref{3} and~\eqref{4} we find that
\begin{equation}
h^0(X, \cO_X(\cC))=70, \quad h^1(X, \cO_X(\cC))=0
\label{1new}
\end{equation}
and, therefore, 
\begin{equation}
n(V)=69.
\label{2.118}
\end{equation}
The number of generations and the five-brane class were computed in {\bf Example 1} in Section 2.

\vspace{0.5cm}
\noindent
{\bf Example 2.} 
Consider the  vector bundle specified by $B={\mathbb F}_1, G=SU(3)$, the 
spectral cover
\begin{equation}
{\cal C}\in |3 \s +\pi^*(8\cS+9\cE)|
\label{2.119}
\end{equation}
and 
\begin{equation}
\lambda=\frac{3}{2}.
\label{2.119.1}
\end{equation}
As discussed previously, the coefficients $a=8, b=9$ do not satisfy~\eqref{2.16} and, 
hence, spectral cover~\eqref{2.119} is not positive.
Using eqs.~\eqref{3} and~\eqref{4}, we find that
\begin{equation}
h^0(X, \cO_X(\cC))=64, \quad h^1(X, \cO_X(\cC))=1.
\label{2new}
\end{equation}
Therefore,
\begin{equation}
n(V)=64.
\label{2.120}
\end{equation}
The number of generations and the five-brane class were computed in {\bf Example 2} in Section 2.

The number of moduli we obtain is smaller, roughly by a factor of two, than in 
the examples presented in~\cite{BDO1}. Thus, by considering bundles with 
non-positive spectral cover, we significantly reduce 
the number of moduli.


\subsection{Bundles with the Minimal Number of Moduli}


Having found expressions for the number of instanton moduli which are valid for all
spectral covers, it is natural to ask under what conditions $n(V)$ is minimal.
First consider $h^0 (X, \cO_X(\cC))$. Let $\cC$ and $\cC^{\prime}$ 
be two irreducible spectral covers. ${\cal C}$ being irreducible means that
in the linear system $|\cC|$ there are irreducible representatives. Also assume that
\begin{equation}
\cC^{\prime} =\cC+ \delta \cC,
\label{2.106}
\end{equation}
where $\delta \cC$ is non-vanishing and effective. 
We want to understand the relation between $h^0 (X, \cO_X(\cC))$ and 
$h^0 (X, \cO_X(\cC^{\prime}))$. The following fact 
is useful: holomorphic sections in $h^0 (X, \cO_X(\cC))$ 
can be thought of as holomorphic sections in $h^0 (X, \cO_X(\cC^{\prime}))$
which vanish along $\delta \cC$. In particular, this says that 
\begin{equation}
h^0 (X, \cO_X(\cC^{\prime})) \geq h^0 (X, \cO_X(\cC)),
\label{2.107} 
\end{equation}
with equality holding if every section of $ \cO_X(\cC^{\prime})$ vanishes along
$\delta \cC$. Equality in~\eqref{2.107} has a simple geometrical interpretation.
It implies that the defining equation for every divisor in $|\cC^{\prime}|$
factorizes, with one factor being the equation of $\delta \cC$. It follows that all divisors
in $|\cC^{\prime}|$ are reducible. This contradicts our assumption that the spectral cover 
$\cC^{\prime}$ is irreducible. Therefore
\begin{equation}
h^0 (X, \cO_X(\cC^{\prime})) > h^0 (X, \cO_X(\cC)).
\label{2.108} 
\end{equation}
Recall from~\eqref{2.61} that, for given $r$ and $n$, the minimal allowed values of the 
the coefficients $a$ and $b$ are
\begin{equation}
a=2n, \quad b=n(r+2).
\label{2.109}
\end{equation}
Denote the associated ``minimal'' spectral cover by $\cC$. It follows 
from~\eqref{2.61} that any other allowed spectral cover is of the form~\eqref{2.106}
where $\delta \cC$ is non-vanishing and effective. Therefore,~\eqref{2.108} holds.
Clearly, then, $h^0 (X, \cO_X(\cC))$ is minimal when the coefficients $a$ and $b$
in the definition of the spectral cover~\eqref{2.60} satisfy~\eqref{2.109}.
It is important to note that neither coefficient $a$ nor $b$ in~\eqref{2.109} 
satisfies the inequalities in~\eqref{2.16}. Hence, the minimal spectral cover is 
never positive. 
It is straightforward to check that for such values of $a$ and $b$ 
$h^1 (X, \cO_X(\cC))$ vanishes and, hence, it is also minimized. Furthermore, note
that for $r=1$ the coefficients $a$ and $b$ satisfy the inequality $b \geq a+n-1$.
Hence, for any $r$ and $n$, the number of moduli in the minimal case is 
of the form ~\eqref{2.90}.
Inserting~\eqref{2.109} into~\eqref{2.90} yields
\begin{equation}
n_{min}(V)=\frac{1}{3}(4n^3+23n-3).
\label{2.110}
\end{equation}
It is easy to show that this expression is indeed an integer.
Eq.~\eqref{2.110} represents the minimal number of moduli for any rank $n$ vector bundle.
Note that it is independent of $r$. 
Let us summarize our results.
\begin{itemize}
\item
For
\begin{equation}
r=0, 1, 2, \quad G=SU(n)
\label{line}
\end{equation}
and 
\begin{equation}
a=2n, \quad b=n(r+2)
\label{doubleline}
\end{equation}
the number of moduli is minimal and given by
\begin{equation}
n_{min}(V)=\frac{1}{3}(4n^3+23n-3).
\label{check}
\end{equation}
\end{itemize}

Let us now consider an example.


\subsection{Example with Minimal Number of Moduli}


{\bf Example 3.} Consider the vector bundle specified by $B={\mathbb F}_r, G=SU(3)$ and the 
spectral cover
\begin{equation}
{\cal C}\in |3 \s +\pi^*(6\cS+3(r+2)\cE)|.
\label{2.112}
\end{equation}
Then $n=3$ and the parameters $a$ and $b$ satisfy eq.~\eqref{2.109}. Therefore, we can use 
eq.~\eqref{check}
to evaluate the number of moduli. We find that
\begin{equation}
n_{min}(V)=58.
\label{2.113}
\end{equation}
This is the minimal number of moduli for a rank $3$ instanton on a Calabi-Yau 
threefold elliptically fibered over any Hirzebruch surface ${\mathbb F}_r$, $r=0,1,2$.
Using eqs.~\eqref{1.9}, \eqref{1.19}, \eqref{1.20} 
and~\eqref{2.112}, we find that the number of generations in this model is
\begin{equation}
N_{gen}=0.
\label{2.114}
\end{equation}
Furthermore, the five-brane class is 
\begin{equation}
[W]=\sigma \cdot \pi^*(18 \cS +27 \cE) +96 F
\label{2.115}
\end{equation}
for every choice of parameter $\lambda$. Note that $[W]$ is effective.\\

Recall that the two examples in the previous subsection each had $n=3$ and $r=1$. 
{\bf Example 1} was specified by
\begin{equation}
a=6, \quad b=10
\label{2.AA}
\end{equation}
and had
\begin{equation}
n(V)=69
\label{2.AAA}
\end{equation}
moduli. Similarly, {\bf Example 2} was specified by
\begin{equation}
a=8, \quad b=9
\label{2.BB}
\end{equation}
and had
\begin{equation}
n(V)=64
\label{2.BBB}
\end{equation}
moduli. Since coefficients $b$ in~\eqref{2.AA} and $a$ in~\eqref{2.BB} do not 
satisfy~\eqref{2.109}, we expect
$n(V)$ in each example to exceed 58, which they do.


\section{Small Instanton Transitions}


\subsection{Transition Moduli}


Consider a vacuum configuration of heterotic M-theory consisting 
of two end-of-the-world branes with vector bundles $V_1$ and $V_2$ and
five-branes wrapped on the holomorphic curve
\begin{equation}
[W]=\s \cdot \pi^*w \ +a_fF.
\label{3.1}
\end{equation}
We assume that the vector bundle $V_2$ is trivial and $V_1$, which we denote by $V$,
is defined by the spectral cover
\begin{equation}
\cC\in |n \s +\pi^* \eta|.
\label{3.2}
\end{equation}
We will always take $\cC$ to be effective and irreducible. 

Consider any 
effective, irreducible curve component of $[W]$. It was shown 
in~\cite{OPP} that one can move this curve to a boundary brane and 
absorb it as a 
small instanton transition into the vector bundle $V$. This results in a new 
bundle $V^{\prime}$. 
A transition which absorbs
only a horizontal component of $[W]$, that is, $\sigma 
\cdot \pi^{*}z$ where $z \subseteq w$, modifies the spectral cover~\eqref{3.2}
to 
\begin{equation}
\cC^{\prime} \in |n \s +\pi^* (\eta+z)|,
\label{3.3}
\end{equation}
which describes a new vector bundle $V^{\prime}$. Transitions of this type
preserve the low-energy gauge group $H$, but change the third Chern class of the 
bundle and, hence, the number of families on the visible brane.
Such transitions are called ``chirality changing''. It was also
demonstrated in~\cite{OPP} these transitions do not alter the topological type of the line 
bundle $\cN$. 
It is also possible to have a transition in which 
a purely vertical component of the five-brane curve gets absorbed~\cite{OPP}.
However, in this case, the new bundle is reducible. The low-energy gauge 
group changes whereas the number of families does not. The number of 
moduli for these bundles was calculated in~\cite{Reduce}. We will not consider 
such transitions in this paper.

An analogous, but simpler, transition takes place in type II string theories in
the $Dp-D(p+4)$ system~\cite{Douglas}.
The $Dp-D(p+4)$ system is describable by a supersymmetric field theory with eight
supercharges. The moduli space of this system consists of two branches, the Coulomb 
branch and the Higgs branch. The Coulomb branch describes positions of the
$Dp$-brane away from the $D(p+4)$-brane. The Higgs branch describes how 
the $Dp$-brane can get dissolved into the $D(p+4)$-brane. Geometrically, the Higgs branch 
is isomorphic to the one-instanton moduli space on a $4$-manifold which is just the ADHM moduli space.
In the heterotic M-theory case, the analog of the Coulomb branch is the moduli space 
of five-branes. The analog of the Higgs branch is the space of transition moduli.
The concept of transition moduli was introduced in~\cite{BDO1}. They are the new moduli 
that arise after the transition and represent a Calabi-Yau threefold analog of the 
ADHM one-instanton moduli space. 
See~\cite{BDO1} for various properties of the transition moduli.

The number 
of the transition moduli can be computed as
\begin{equation}
n_{tm}=n(V^{\prime})-n(V),
\label{3.4}
\end{equation}
where $n(V)$ is given by eq.~\eqref{2.90} or eq.~\eqref{2.91}.
Transition moduli appear to play an important role in heterotic compactifications.
In~\cite{BDO2, BDO3}, the non-perturbative superpotentials for vector bundle moduli
were studied. It was shown that the superpotential due to a string wrapped on an 
isolated curve $\s \cdot \pi^* \cS$ in a Calabi-Yau manifold fibered over 
${\mathbb F}_r$ is proportional to a polynomial of the transition moduli associated with
the curve $\s \cdot \pi^* \cS$. 
For example, in the case
\begin{eqnarray}
&&r=2, \quad n=3, \quad \lambda =\frac{3}{2},\nonumber \\
&&{\cal C}\in |3 \s +\pi^*(a \cS+b\cE)|
\label{3.5}
\end{eqnarray}
with
\begin{equation}
a>6, \quad b-2a=2,
\label{3.6}
\end{equation}
the non-perturbative superpotential due to a string wrapped on the curve
$\s \cdot \pi^* \cS$ is given by
\begin{equation}
W \propto {\cal R}^4,
\label{3.7}
\end{equation}
where 
\begin{equation}
{\cal R}=\a_1 \b_2 \g_3 -\a_1 \b_3 \g_2 +\a_2 \b_3 \g_1 -\a_2 \b_1 \g_3+
\a_3 \b_1 \g_2-\a_3 \b_2 \g_1.
\label{3.8}
\end{equation}
Here $\a_i, \b_j, \g_k$ are nine transition moduli associated with the curve
$\s \cdot \pi^* \cS$. See~\cite{BDO3} for details. 
However, all results and applications of transition moduli 
in~\cite{BDO1, BDO2, BDO3} hold only for vector bundles with 
positive spectral cover. In the next subsection, we 
will define, and give a geometric meaning to, the transition moduli
in the case of arbitrary, irreducible spectral covers. 


\subsection{Geometric Interpretation of Transition 
Moduli for Arbitrary Irreducible Spectral Covers}


The number of 
instanton moduli was given in~\eqref{2.14} by
\begin{equation}
n(V)=(h^0(X, {\cal O}_X({\cal C}))-1)+h^1(X, {\cal O}_X({\cal C})),
\label{3.2.1}
\end{equation}
where $h^0(X, {\cal O}_X({\cal C}))-1$ counts the parameters of the 
spectral cover $\cC$ and $h^1(X, {\cal O}_X({\cal C}))$ 
the moduli of the line bundle $\cN$.
The number of transition 
moduli is
\begin{eqnarray}
&&n_{tm}=n(V^{\prime}) -n(V) =\nonumber \\
&&[h^0(X, {\cal O}_X({\cal C^{\prime}})) -h^0(X, {\cal O}_X({\cal C}))]+
[h^1(X, {\cal O}_X({\cal C^{\prime}})) -h^1(X, {\cal O}_X({\cal C}))].
\label{3.2.2}
\end{eqnarray}
In this subsection, we want to find a geometric 
interpretation for eq.~\eqref{3.2.2}.
This can be important for calculating 
non-perturbative superpotentials, as was the case for positive 
spectral covers in refs.~\cite{BDO2, BDO3}.
First, consider the situation when
\begin{equation}
h^1(X, {\cal O}_X({\cal C^{\prime}})) =h^1(X, {\cal O}_X({\cal C}))=0.
\label{3.2.3}
\end{equation}
Then
\begin{equation}
n_{tm}=h^0(X, {\cal O}_X({\cal C^{\prime}})) -h^0(X, {\cal O}_X({\cal C})),
\label{3.2.4}
\end{equation}
which coincides with the number of transition 
moduli of bundles with positive spectral cover~\cite{BDO1}.
Hence, this difference has the same interpretation as in~\cite{BDO1},
namely, that the difference in~\eqref{3.2.4} is related to the moduli
of the spectral cover restricted to the lift of the transition curve.
More precisely, if the linear systems $|\cC^{\prime}|$ and $|\cC|$ are related as
\begin{equation}
|\cC^{\prime}|= |\cC+\pi^* z|
\label{3.2.5}
\end{equation}
for some effective curve $z$, then~\cite{BDO1}
\begin{equation}
n_{tm}=h^0(\pi^*z, \cO_X(\cC^{\prime})|_{\pi^* z}).
\label{3.2.6}
\end{equation}
To prove this, consider the short exact sequence
\begin{equation}
0 \rightarrow \cO_X(\cC)  \rightarrow \cO_X(\cC^{\prime})
\rightarrow \cO_{\pi^*z}(\cC^{\prime}|_{\pi^* z}) \rightarrow 0.
\label{3.2.7}
\end{equation}
The associated cohomology sequence is
\begin{eqnarray}
&& 0 \rightarrow H^{0}(X, \cO_{X}(\cC)) \rightarrow H^{0}(X, \cO_{X}(\cC^{\prime})) 
\rightarrow H^0(\pi^* z, \cO_{\pi^*z}(\cC^{\prime}|_{\pi^* z})) \nonumber \\
&& \stackrel{M}{\rightarrow}
H^{1}(X, \cO_{X}(\cC)) \stackrel{\a}{\rightarrow} H^{1}(X, \cO_{X}(\cC^{\prime})) 
\rightarrow H^1(\pi^* z, \cO_{\pi^*z}(\cC^{\prime}|_{\pi^* z}))
\rightarrow \dots .
\label{3.2.8}
\end{eqnarray}
When eq.~\eqref{3.2.3} holds, sequence~\eqref{3.2.8} simplifies to 
\begin{equation}
0 \rightarrow H^{0}(X, \cO_{X}(\cC)) \rightarrow H^{0}(X, \cO_{X}(\cC^{\prime})) 
\rightarrow H^0(\pi^* z, \cO_{\pi^*z}(\cC^{\prime}|_{\pi^* z}))\rightarrow 0
\label{3.2.9}
\end{equation}
and it follows  immediately  that 
\begin{equation}
h^0(X, {\cal O}_X({\cal C^{\prime}})) -h^0(X, {\cal O}_X({\cal C}))=
h^0(\pi^*z, \cO_X(\cC^{\prime})|_{\pi^* z}).
\label{3.2.10}
\end{equation}
This proves eq.~\eqref{3.2.6}. It particular, it follows from \eqref{3.2.10} that
\begin{equation}
{\mathbb P}H^0(X,{\mathcal O}_X(C)) \subset {\mathbb P}H^0(X,{\mathcal O}_X(C^{'})).
\label{r3}
\end{equation}
That is, we can explicitly identify the moduli space of the vector bundle $V$ with a subspace of 
the moduli of bundle $V^{'}$. The normal directions to this subspace are simply given by 
global sections in $\cO_{\pi^*z}(\cC^{\prime}|_{\pi^* z})$.

Let us consider the general case when eq.~\eqref{3.2.3} is no 
longer valid. 
The interpretation of transition moduli is now much more subtle.
Again consider sequence~\eqref{3.2.8}
and the map $\alpha$. 
First, recall from \eqref{r1} that
\begin{equation}
H^1({\cal C},{\cal O}_{{\cal C}})=H^1(X, {\cal O}_X({\cal C}))^{*}.
\label{r2}
\end{equation}
An analogous condition holds if we replace ${\cal C}$ with
$\cC^{\prime}$. The  map $\a$ in eq.~\eqref{3.2.8} is 
\begin{equation}
\a : H^{1}(X, \cO_{X}(\cC)) \rightarrow H^{1}(X, \cO_{X}(\cC^{\prime})).
\label{3.2.15}
\end{equation}
Dualizing $\a$ and using eq.~\eqref{r2}, we get 
\begin{equation}
\a^* : H^{1}(\cC^{\prime}, \cO_{\cC^{\prime}}) \rightarrow H^{1}(\cC, \cO_{\cC}).
\label{3.2.16}
\end{equation}
Now $H^{1}(\cC, \cO_{\cC})$, up to a discrete lattice, is 
the moduli space of line bundles ${\cal N}$ on ${\cal C}$.
Therefore, for fixed $\cC$ and $\cC^{'}$, $\a$ gives an explicit map between line bundles on $\cC$ and $\cC^{'}$.
If we consider for a moment spectral covers $\cC^{'}$ which are in the image of the inclusion \eqref{r3}, than 
$\cC$ is contained in $\cC^{\prime}$ and it is natural to interpret the map $\alpha^*$ as the 
restriction map. That is, it takes a line bundle on $\cC^{\prime}$ and 
restricts it to $\cC$. 

At this point, it is useful to recall some  linear algebra from Appendix B.
Let $f$ be a linear map between two linear spaces ${\cal A}$ and ${\cal B}$.
If we denote by $f^{*}$ the dual map, it follows from Appendix B that
\begin{eqnarray}
&&dim{\ }ker{\ }f^* = dim{\ }coker{\ }f, \nonumber \\
&&dim{\ }ker{\ }f = dim{\ }coker{\ }f^*.
\label{3.2.20}
\end{eqnarray}
Let us use this fact to interpret the transition moduli.
Sequence~\eqref{3.2.8} can be split as follows
\begin{equation}
0 \rightarrow H^{0}(X, \cO_{X}(\cC)) \rightarrow H^{0}(X, \cO_{X}(\cC^{\prime})) 
\rightarrow ker{\ }M \rightarrow 0
\label{split} 
\end{equation}
and
\begin{equation}
0 \rightarrow im{\ }M \rightarrow
H^{1}(X, \cO_{X}(\cC)) \stackrel{\a}{\rightarrow} H^{1}(X, \cO_{X}(\cC^{\prime})) 
\rightarrow coker{\ }\a \rightarrow 0.
\label{3.2.21}
\end{equation}
The first of these sequences states that $ker{\ }M$ are the moduli of the spectral cover
${\cal C}^{\prime}$ which are not available for $\cC$. That is, they are the new moduli
that arise after the transition and modify the spectral cover. 
In other words, $dim {\ }ker{\ }M$ simply counts how many   extra moduli the 
spectral cover ${\cal C}^{\prime}$ has comparing to $\cC$.
Obviously,
\begin{equation}
dim{\ }ker{\ }M =h^{0}(X, \cO_{X}(\cC^{\prime})) -h^{0}(X, \cO_{X}(\cC)).
\label{3.2.22}
\end{equation}
We also need to understand $im{\ }M$. Since the sequence~\eqref{3.2.8} is exact,
it follows that
\begin{equation}
im{\ }M =ker{\ }\a.
\label{3.2.23}
\end{equation}
Furthermore, from eqs.~\eqref{3.2.20} we get
\begin{equation}
im{\ }M=coker{\ }\a^*.
\label{3.2.24}
\end{equation}
As discussed previously, the map $\a^*$ relates  line
bundles on ${\cal C}^{\prime}$ to $\cC$. Then $coker{\ }\a^*$, by definition,
are those line bundle on $\cC$ which cannot be related to  line bundles on ${\cal C}^{\prime}$. 
The last space in~\eqref{3.2.21} to interpret is $coker{\ }\a$.
Since, by eq.~\eqref{3.2.20}, 
\begin{equation}
coker{\ }\a =ker{\ }\a^*,
\label{3.2.25}
\end{equation}
$coker{\ }\a$ are just those non-trivial line bundles on ${\cal C}^{\prime}$
which correspond to trivial  line bundles on $\cC$.
By definition, transition moduli are new moduli that arise after the small 
instanton transition. Therefore, to find the number of transition moduli we 
have to add the parameters of the spectral cover available for 
${\cal C}^{\prime}$ but not for $\cC$ (that is, $ker{\ }M$) to the 
parameters of those line bundles which are non-trivial on 
${\cal C}^{\prime}$ but trivial on $\pi^{*}z$ (that is,
$coker{\ }\a$). However, there is a caveat. As established, 
$im{\ }M$ is the set of those line bundles which exist on $\cC$ 
but cannot be related to line bundles  from ${\cal C}^{\prime}$.
If $im{\ }M$ is non-zero, such moduli will disappear after the 
transition. Therefore, the number of transition moduli must be 
equal to
\begin{equation}
n_{tm}=dim{\ }ker{\ }M +dim{\ }coker{\ }\a-dim{\ }im{\ }M.
\label{3.2.26}
\end{equation}
This equation counts how many new moduli appear after 
the small instanton transition. Let us see whether 
eq~\eqref{3.2.26} is equivalent to eq.~\eqref{3.2.2}.
From the sequences~\eqref{split} and~\eqref{3.2.21}, it follows that
\begin{equation}
dim{\ }ker{\ }M =h^{0}(X, \cO_{X}(\cC^{\prime})) -h^{0}(X, \cO_{X}(\cC))
\label{3.2.27}
\end{equation}
and 
\begin{equation}
coker{\ }\a -im{\ }M=
h^{1}(X, \cO_{X}(\cC^{\prime})) -h^{1}(X, \cO_{X}(\cC)).
\label{3.2.28}
\end{equation}
Substituting eqs.~\eqref{3.2.27} and~\eqref{3.2.28} into eq.~\eqref{3.2.26},
we obtain precisely eq.~\eqref{3.2.2}.

Let us now summarize our interpretation of transition moduli
for arbitrary spectral covers. 
\begin{itemize}
\item 
Some transition moduli arise as new parameters of the spectral cover.
Note that, in general, they do not correspond to the parameters of the spectral cover 
restricted to the lift of the curve
$H^0(\pi^*z, \cO_{\pi^* z}(\cC^{\prime}|_{\pi^* z}))$.
They do so only if the map $\a$ is an isomorphism, as follows from the 
sequence~\eqref{3.2.8}. In this case, the map $M$ is the zero map
and its kernel is the whole space 
$H^0(\pi^*z, \cO_{\pi^* z}(\cC^{\prime}|_{\pi^* z}))$. In particular, this 
happens when the spectral covers $\cC^{\prime}$ and $\cC$ are both positive.

\item 
Some transition moduli arise as parameters of the line bundle 
$\cN$ on $\cC^{\prime}$ which correspond to the trivial 
line bundle on $\cC$. 

\item
Some moduli can disappear in the process of the transition.
They correspond to moduli of those line bundles on $\cC$ which 
are not related to  line bundles on $\cC^{\prime}$.
\end{itemize}

Two further remarks are in order. First, we note that during a small instanton transition
the number of parameters of the spectral cover must
always increase. Recall from Subsection 3.5. that
the spectral covers before and after the transition, $\cC$ and $\cC^{\prime}$ respectively,
are related by 
\begin{equation}
\cC^{\prime}=\cC+\delta \cC,
\label{3.34}
\end{equation}
where $\cC$, $\cC^{\prime}$ and $\delta \cC$ are all effective and irreducible and, hence,
that
\begin{equation}
h^0(X, \cO_X(\cC^{\prime})) >h^0(X, \cO_X(\cC)).
\label{3.35}
\end{equation}
Therefore, the number of parameters of the spectral cover always becomes larger.
Second, we note that this is not true for line bundle moduli.
In the next subsection, we will give explicit examples of 
small instanton transition where the number of line bundle moduli
decreases. This indicates that the moduli described in the 
last item above  do indeed exist and, generically, that $im{\ }M$ does not vanish.


\subsection{Examples}


In this concluding subsection, we give several examples of 
chirality changing small instanton transitions
involving various vacua with three generations. We will observe
that the number of generations changes after the transition.

\vspace{0.5cm}
\noindent
{\bf Example A.}  In this example, we choose 
$B={\mathbb F}_1, G=SU(3), a=6, b=9, \lambda=\frac{1}{2}$.
Note that this case corresponds to {\bf Example 3} in Section 3 with $r=1$.
From eqs.~\eqref{1.9}, \eqref{1.19}, \eqref{1.20} we get
\begin{equation}
N_{gen}=0
\label{3.9}
\end{equation}
and 
\begin{equation}
[W]=\s \cdot \pi^*
(18 \cS +27 \cE) +96F.
\label{3.10}
\end{equation}
From eqs.~\eqref{1},~\eqref{2} and~\eqref{2.90} we obtain
\begin{equation}
h^0 (X, \cO_X(\cC)) =59, \quad h^1 (X, \cO_X(\cC)) =0
\label{3new}
\end{equation}
and
\begin{equation}
n(V)=58.
\label{3.11}
\end{equation}

\noindent {\bf A1.} Let us make a transition with the curve 
$\s \cdot \pi^* \cE$. The remaining five-brane class is
\begin{equation}
[W^{\prime}]=
\s \cdot \pi^*
(18 \cS +26 \cE) +96F,
\label{3.12}
\end{equation}
while the new spectral cover is given by 
\begin{equation}
a^{\prime}=6, \quad b^{\prime}=10.
\label{3.13}
\end{equation}
This corresponds to {\bf Example 1} in Section 2.
From eqs.~\eqref{1},~\eqref{2} and~\eqref{2.90} we obtain
\begin{equation}
h^0 (X, \cO_X(\cC^{\prime})) =70, \quad h^1 (X, \cO_X(\cC^{\prime})) =0
\label{4new}
\end{equation}
and
\begin{equation}
n(V)=69.
\label{3.14}
\end{equation}
The number of transition moduli can be calculated by using eq.~\eqref{3.14}
\begin{equation}
n_{tm}=n(V^{\prime})-n(V) =11.
\label{3.15}
\end{equation}
The number of generations was computed in~\eqref{1.30} and is given by
\begin{equation}
N_{gen}=3.
\label{3.16}
\end{equation}
This is an example of a pure spectral cover transition. 
It changes the number of generations from zero to three.\\

\noindent {\bf A2.} Now consider a transition with the curve 
$\s \cdot \pi^* \cS$. The resulting five-brane class is 
\begin{equation}
[W^{\prime \prime}]=\s \cdot \pi^*
(17 \cS +27 \cE) +96F.
\label{3.17}
\end{equation}
and the new spectral cover is characterized by 
\begin{equation}
a^{\prime \prime}=7, \quad b^{\prime \prime}=9.
\label{3.18}
\end{equation}
From eqs.~\eqref{1},\eqref{2} and~\eqref{2.90} we obtain
\begin{equation}
h^0 (X, \cO_X(\cC^{\prime \prime})) =63, \quad h^0 (X, \cO_X(\cC^{\prime \prime})) =0
\label{5new}
\end{equation}
and
\begin{equation}
n(V)=62.
\label{3.19}
\end{equation}
The number of transition moduli can be calculated by using eq.~\eqref{3.14}
\begin{equation}
n_{tm}=n(V^{\prime \prime})-n(V) =4.
\label{3.20}
\end{equation}
From eq.~\eqref{1.20} we find that
\begin{equation}
N_{gen}=1.
\label{3.21}
\end{equation}
This is also an example of a pure spectral cover transition. 
It changes the number of generations from zero to one.

\vspace{0.5cm}
\noindent
{\bf Example B.} In this example, we choose 
$B={\mathbb F}_1, G=SU(3), a=8, b=9, \lambda=\frac{3}{2}$. 
This corresponds to {\bf Example 2} in Section 2,
where the number of generations and the five-brane class were computed 
and found to be
\begin{equation}
N_{gen}=3
\label{3.23}
\end{equation}
and 
\begin{equation}
[W]=
\s \cdot \pi^*
(16 \cS +27 \cE) +90F.
\label{3.23.1}
\end{equation}
From eqs.~\eqref{3},~\eqref{4} and~\eqref{2.91} we obtain
\begin{equation}
h^0 (X, \cO_X(\cC)) =64, \quad h^1 (X, \cO_X(\cC)) =1
\label{6new}
\end{equation}
and
\begin{equation}
n(V)=64.
\label{3.24}
\end{equation}

\noindent {\bf B1.} Let us make a transition with the curve 
$\s \cdot \pi^* \cE$. The resulting five-brane class is 
\begin{equation}
[W^{\prime}]=\s \cdot \pi^*
(16 \cS +26 \cE) +90F.
\label{3.24.1}
\end{equation}
The new spectral cover is characterized by 
\begin{equation}
a^{\prime}=8, \quad b^{\prime}=10.
\label{3.25}
\end{equation}
From eqs.~\eqref1{},~\eqref{2} and~\eqref{2.90} we obtain
\begin{equation}
h^0 (X, \cO_X(\cC^{\prime})) =81, \quad h^1 (X, \cO_X(\cC^{\prime})) =0
\label{7new}
\end{equation}
and
\begin{equation}
n(V)=80.
\label{3.26}
\end{equation}
The number of transition moduli can be calculated by using eq.~\eqref{3.14}
\begin{equation}
n_{tm}=n(V^{\prime})-n(V) =16.
\label{3.27}
\end{equation}
From eq.~\eqref{1.20}, we find that the number of generations is  
\begin{equation}
N_{gen}=18.
\label{3.28}
\end{equation}
Note that, as always, the number of spectral cover moduli has increased.
However, this is a clear example of a
transition where the number of line bundle moduli decreases.
\\

\noindent {\bf B2.} Now consider a transition with the curve 
$\s \cdot \pi^* \cS$. The resulting five-brane class is 
\begin{equation}
[W^{\prime \prime}]=
\s \cdot \pi^*
(15 \cS +27 \cE) +90F.
\label{3.29}
\end{equation}
and the new spectral cover is characterized by 
\begin{equation}
a^{\prime \prime}=9, \quad b^{\prime \prime}=9.
\label{3.30}
\end{equation}
From eqs.~\eqref{3},~\eqref{4} and~\eqref{2.91} we obtain
\begin{equation}
h^0 (X, \cO_X(\cC^{\prime \prime})) =66, \quad h^1 (X, \cO_X(\cC^{\prime \prime})) =4
\label{8new}
\end{equation}
and
\begin{equation}
n(V)=69.
\label{3.31}
\end{equation}
The number of transition moduli can be calculated by using eq.~\eqref{3.14}
\begin{equation}
n_{tm}=n(V^{\prime \prime})-n(V) =5.
\label{3.32}
\end{equation}
From eq.~\eqref{1.20} we find that
\begin{equation}
N_{gen}=0.
\label{3.33}
\end{equation}
Note that, in this example, the number of both the spectral cover 
and the line bundles parameters increases.


\section{Conclusion}


In this paper, we have given a complete classification of 
the number of instanton moduli for Calabi-Yau threefolds
elliptically fibered over the Hirzebruch surfaces ${\mathbb F}_r, r=0, 1, 2$.
The results are presented in eqs.~\eqref{2.90} and~\eqref{2.91}.
An expression for the minimal possible 
number of moduli for a vector bundle of fixed $r$ and $n$ was presented. It is independent of $r$
and given by
\begin{equation}
n_{min}(V)=\frac{1}{3}(4n^3+23n-3).
\label{c1}
\end{equation}
For example, for $n=3$ the minimal number of vector bundle moduli is $58$.
Hence, in GUT theories built on such vacua,
the number of instanton moduli is always rather large. 
It seems unlikely that this number can substantially be reduced
by considering instantons on Calabi-Yau threefolds fibered 
over different bases. However, it is natural to expect 
that the number of moduli will be greatly reduced for
standard model-like bundles on Calabi-Yau threefolds with
non-trivial discrete homotopy group~\cite{Standard1, Standard2, FF, Rene1, Rene2, SU(4), Volker}. 
Such bundles are more rigid and one does not expect many moduli.
We also gave an interpretation for transition moduli. 
Such moduli arise after the small instanton transition 
and represent a Calabi-Yau threefold generalization
of the ADHM one-instanton moduli space. We saw that the number 
of transition moduli in certain cases can be reduced comparing to 
previous results in~\cite{BDO1} and can be as few as four. 
We supplemented all of these concepts with concrete examples.


\section{Appendix A: Consistency Checks}


In this Appendix, we present some consistency
checks of eqs.~\eqref{2.90} and~\eqref{2.91}. First, note that
the positivity conditions~\eqref{2.16}
are a special case of the conditions stated in eq.~\eqref{2.89.1}.
Therefore,
eq.~\eqref{2.90} should be valid for positive spectral covers.
This is indeed correct as eq.~\eqref{2.90} coincides with eq.~\eqref{2.19}.
The second consistency check we can perform is the following.
Comparing eqs.~\eqref{2.90} and~\eqref{2.91} with the expression for the Euler 
characteristic~\eqref{2.18.1}, we see that eq.~\eqref{2.90}
can be written as
\begin{equation}
n(V)=\chi_E(X, \cO_X(\cC)) -1,
\label{2.92}
\end{equation}
whereas eq.~\eqref{2.91} can be expressed as
\begin{equation}
n(V)=\chi_E(X, \cO_X(\cC)) -1 +2h^1((X, \cO_X(\cC)).
\label{2.93}
\end{equation}
This indicates that 
\begin{equation}
H^2(X, \cO_X(\cC))=H^3(X, \cO_X(\cC))=0
\label{2.94}
\end{equation}
for all spectral covers (of course, we always assume that conditions~\eqref{1.11}-~\eqref{1.13} are 
satisfied) as mentioned in eqs.~\eqref{2.74.1} and~\eqref{2.88.1} above . 
Let us prove eq.~\eqref{2.94} directly, using independent methods. 
Start with $H^2(X, \cO_X(\cC))$. From Leray spectral sequences, it follows that
\begin{equation}
H^2(X, \cO_X(\cC)) =H^2(B,\pi_* \cO_X(\cC)).
\label{2.95}
\end{equation}
We have used the facts that
\begin{equation}
H^0(B, R^2\pi_* \cO_X(\cC))=0,
\label{2.96}
\end{equation}
since the fiber of the projection $\pi$ is one-dimensional
and
\begin{equation}
H^1(B, R^1\pi_* \cO_X(\cC))=0,
\label{2.97}
\end{equation}
as was proven previously in eq~\eqref{2.40}.
Next, we relate $H^2(X, \cO_X(\cC))$ to 
$H^2({\cal S}, \rho_* \pi_* \cO_X(\cC))$. 
Note that 
\begin{equation}
H^0(\cS, R^2\rho_* \pi_* \cO_X(\cC))=0
\label{2.99}
\end{equation}
since the fiber of the projection $\rho$, the curve $\cE$, is one-dimensional
and 
\begin{equation}
H^1(\cS, R^1\rho_* \pi_* \cO_X(\cC))=0,
\label{2.100}
\end{equation}
as shown previously (see eq.~\eqref{2.48} and discussion just above it).
Using this and a Leray spectral sequence, we find that 
\begin{equation}
H^2(B,\pi_* \cO_X(\cC))=H^2(\cS, \rho_* \pi_* \cO_X(\cC)).
\label{2.98}
\end{equation}
This vanishes since $\cS$ is one-dimensional. 
Combining this result with eq.~\eqref{2.95} proves that
\begin{equation}
H^2(X, \cO_X(\cC)) =0.
\label{2.101}
\end{equation}
Similarly, we can prove that $H^3(X, \cO_X(\cC))$
vanishes. 
To see this, note that
\begin{equation}
H^2(B, R^1\pi_* \cO_X(\cC))=0
\label{2.103}
\end{equation}
since the sheaf $R^1\pi_* \cO(\cC)$ vanishes (see eq.~\eqref{2.40}) and 
\begin{equation}
H^1(B, R^2\pi_* \cO_X(\cC))=H^0(B, R^3\pi_* \cO_X(\cC))=0
\label{2.104}
\end{equation}
since the fiber of $\pi$ is one-dimensional.
Leray spectral sequences then imply that
\begin{equation}
H^3(X, \cO_X(\cC)) =H^3(B,\pi_* \cO_X(\cC)).
\label{2.102}
\end{equation}
But
\begin{equation}
H^3(B, \pi_* \cO_X(\cC))=0
\label{2.105}
\end{equation}
since $B$ is a surface. This proves that 
\begin{equation}
H^3(X, \cO_X(\cC)) =0.
\label{2.105.1}
\end{equation}
Hence, eqs.~\eqref{2.90} and~\eqref{2.91} are indeed consistent 
with eqs.~\eqref{2.92} and~\eqref{2.93}.


\section{Appendix B: Some Linear Algebra}


Let $f$ be a map between two linear spaces ${\cal A}$ and ${\cal B}$. Then, it is 
known that there exists the following short exact sequence
\begin{equation}
0\rightarrow ker{\ }f \rightarrow {\cal A} 
\stackrel{f}{\rightarrow}{\cal B} \rightarrow coker{\ }f \rightarrow 0.
\label{3.2.17}
\end{equation}
Dualizing this sequence, we get
\begin{equation}
0\rightarrow (coker{\ }f)^* \rightarrow {\cal B}^*
\stackrel{f^*}{\rightarrow}{\cal A}^* \rightarrow (ker{\ }f)^* \rightarrow 0.
\label{3.2.18}
\end{equation}
On the other hand, replacing $f$ by $f^*$, ${\cal A}$ 
by ${\cal B}^*$ and ${\cal B}$ by ${\cal A}^*$ in~\eqref{3.2.17} we obtain
\begin{equation}
0\rightarrow ker{\ }f^* \rightarrow {\cal B}^*
\stackrel{f^*}{\rightarrow}{\cal A}^* \rightarrow coker{\ }f^* \rightarrow 0.
\label{3.2.19}
\end{equation}
Comparing eqs.~\eqref{3.2.18} and~\eqref{3.2.19} we conclude that
\begin{eqnarray}
&&dim{\ }ker{\ }f^* = dim{\ }coker{\ }f, \nonumber \\
&&dim{\ }ker{\ }f = dim{\ }coker{\ }f^*.
\end{eqnarray}


\section{Acknowledgment}


The work of E.~I.~B. is supported by NSF grant PHY-0070928. The work of B.~A.~O. is supported 
in part by the Department of Physics and the Math/Physics Research group
at the University of Pennsylvania under cooperative research agreement
No. DE-FG02-95ER40893 with the U.~S. Department of Energy and an NSF
Focused Research Grant DMS0139799 for ``The Geometry of Superstrings''. 
The work of R.~R. is supported by DMS-0244464.



\end{document}